# LOW ENERGY THEOREMS AND THE UNITARITY BOUNDS IN THE EXTRA U(1) SUPERSTRING INSPIRED $E_6$ MODELS

N.K. Sharma[1], Pranav Saxena[2], Prachi Parashar[3*], Ashok K. Nagawat[4] and Sardar Singh[5]

*Theoretical High Energy Physics Group*
*Department of Physics, University of Rajasthan, Jaipur – 302 004 (INDIA).*
**Dept. of Physics & Astronomy, Oklahoma University, Oklahoma, Norman 73019 (USA)*

## ABSTRACT

The conventional method using "Low Energy Theorems" derived by Chanowitz *et al.* [3] does not seem to lead to explicit unitarity limit in the scattering processes of longitudinally polarized gauge bosons for the high energy case in the extra $U(1)$ superstring inspired models, commonly known as $\eta$ model, emanating from $E_6$ group of superstring theory. We have made use of an alternative procedure given by Durand & Lopez [14], which is applicable to supersymmetric grand unified Theories. Explicit unitarity bound on the superpotential couplings (identified as Yukawa couplings) are obtained from both using unitarity constraints as well as using RGE analysis at one loop level utilizing critical couplings concept implying divergence of scalar coupling at $M_G$. These are found to be consistent with finiteness over the entire range $M_Z \leq \sqrt{s} \leq M_G$ i.e. from grand unification scale to weak scale. For completeness the similar approach has been made use of in other models i.e. - $\chi, \psi,$ and $\nu$ models emanating from $E_6$ and it has been noticed that at weak

1 nksharma@uniraj.ernet.in
2 pranav@uniraj.ernet.in
3 prachi@nhn.ou.edu
4 nagawat@uniraj.ernet.in
5 ssingh@uniraj.ernet.in

scale, the unitarity bounds on Yukawa couplings do not differ among $E_6$ extra $U(1)$ models significantly except for the case of $\chi$ model in ***16*** representations.

For the case of $E_6$-$\eta$ model ($\beta_E \cong 9.64$), the analysis using the unitarity constraints leads to the following bounds on various parameters: $\lambda_{t\max.}(M_Z)= 1.294$, $\lambda_{b\max.}(M_Z)= 1.278$, $\lambda_{H\max.}(M_Z)= 0.955$, $\lambda_{D\max.}(M_Z)= 1.312$. The analytical analysis of RGEs at one loop level provides following critical bounds on superpotential couplings i.e. $\lambda_{t,c} \cong 1.295$, $\lambda_{b,c} \cong 1.279$, $\lambda_{H,c} \cong 0.968$, $\lambda_{D,c} \cong 1.315$. Thus superpotential coupling values obtained by both the approaches are in good agreement. Theoretically we have obtained bounds on physical mass parameters using the unitarity constrained superpotential couplings. The bounds are as follows: (i) Absolute upper bound on top quark mass $m_t \le 225\ GeV$, (ii) the upper bounds on the lightest neutral Higgs mass at the tree level is $m_{H_2^0}^{tree} \le 169\ GeV$, and after the inclusion of one loop radiative correction it is $m_{H_2^0} \le 229\ GeV$ when $\lambda_t \neq \lambda_b$ at GUT scale. On the other hand, these are $m_{H_2^0}^{tree} \le 159\ GeV$, $m_{H_2^0} \le 222\ GeV$, respectively, when $\lambda_t = \lambda_b$ at GUT scale. A plausible range for D-quark mass as a function $M_{Z_2}$ is $m_D \approx O\,(3\,TeV)$ for $m_{Z_2} \approx O\,(1\,TeV)$ for the favored values of $\tan\beta \le 1$. The bounds on aforesaid physical parameters in the case of $\chi, \psi,$ and $\nu$ models in the ***27*** representations are almost identical with those of $\eta$ model and are consistent with the present day experimental precision measurements

## I. INTRODUCTION

The past two decades have witnessed a good deal of activity in the field of electroweak symmetry breaking (EWSB) mechanism [1] and the unitarity bound [2,3]. Although $SU(2)_L \times U(1)_Y$ gauge symmetry of electroweak interactions is very successful till date, yet we know nothing about the mechanism that breaks this symmetry and makes $W$ and $Z$ massive. In particular, we are not yet aware whether the physics of electroweak symmetry breaking is weakly or strongly interacting. For the weakly interacting symmetry breaking sector, there are models with elementary Higgs boson [4] which are much lighter than 1 $TeV$. These also include supersymmetric models in which the scale of supersymmetry breaking is of the order of 100 $GeV$. For the strongly interacting symmetry breaking sector, these models allow minimum Higgs [4] mass to be greater than 1 $TeV$. Furthermore, Technicolor models [5] with technihadron spectrum of the order of 1 $TeV$ or above involve breaking through dynamical symmetry breaking mechanism. Due to lack of experimental confirmation, it is yet not possible to arrive at a unique acceptable framework for EWSB. However, it is known that the vacuum expectation value of whatever that breaks the electroweak symmetry is fixed by the Fermi scale [6] through the relation

$$\upsilon_0 = \left(\sqrt{8}G_F\right)^{-\frac{1}{2}} \cong 174\ GeV \cdot \qquad (1)$$

It is also known [3] that the typical mass scale of the symmetry breaking sector ($M_{SB}$) and the strength of the interaction ($\lambda_{SB}$) are correlated such that for $\frac{M_{SB}}{\upsilon_0} \leq 1$, $\lambda_{SB}$ is expected to be very small and may be amenable to perturbative analysis. On the other hand, for $\frac{M_{SB}}{\upsilon_0} \geq 1$, $\lambda_{SB}$ will be very large and the symmetry breaking interaction is expected to be strong. Its analysis would require nonperturbative methods. A useful method for dealing with the latter case is the use of low energy theorems [7], which are useful in the study of scattering of strongly interacting longitudinally, polarized W and Z bosons [3]. These are derived following an analogy of investigating $\pi^{+}\pi^{-} \to \pi^{0}\pi^{0}$ scattering and other similar scattering processes first used by Weinberg [7]. The scattering amplitudes at low energy are completely determined by Eq. (1) along with knowledge of $\rho$ parameter defined by

$$\rho = \left( \frac{M_W}{M_Z \cos\theta} \right)^2 \tag{2}$$

where experimentally $\rho$ =1, within a few percent. This implies the low energy scattering amplitudes for all experimentally viable models of the symmetry breaking sector will have universal values with the mass spectra fully at 1 *TeV* or above. If the spectrum contains bosons lighter that 1 *TeV*, the low energy amplitudes will need modifications.

Chanowitz *et al.* [3] derived the following low-energy theorems for the case of longitudinally polarized gauge bosons scattering amplitudes for the energy range $M_W << \sqrt{s} << \Lambda_{SB}$, in the standard model (SM), where

$$\Lambda_{SB} = \min\{M_{SB}, 4\pi\upsilon_0\}. \tag{3}$$

The theorems are expressed as:

$$M\left(W_L^+ W_L^- \to Z_L Z_L\right) \approx \frac{g^2 s}{4M_W^2} = \frac{s}{\upsilon_0^2}, \tag{4}$$

$$M\left(W_L^+ W_L^- \to Z_L Z_L\right) \approx -\frac{g^2 u}{4M_W^2} = -\frac{u}{\upsilon_0^2}, \tag{5}$$

$$M\left(W_L^+ W_L^- \to Z_L Z_L\right) \approx 0, \tag{6}$$

$$M\left(W_L^+ W_L^- \to Z_L Z_L\right) \approx \frac{g^2 t}{4M_W^2} = \frac{t}{\upsilon_0^2}, \tag{7}$$

$$M\left(W_L^+ W_L^+ \to W_L^+ W_L^+\right) \approx M\left(W_L^- W_L^- \to W_L^- W_L^-\right) \approx \frac{g^2 s}{4M_W^2} = \frac{s}{v_0^2}, \qquad (8)$$

where $M_W = \frac{1}{2} g v_0$, and $s$, $u$, $t$ are Mandelstam variables. For establishing these theorems three different approaches have been used [3]: (i) a power counting analysis carried out in the unitarity (U) gauge, (ii) a current algebra derivation in the Landau gauge along with the use of equivalence theorem, and (iii) the effective chiral lagrangian method utilizing unitarity or renormalization gauges.

In the discussion of low energy theorems and unitarity using perturbative analysis the main argument consists in dividing the U-gauge scattering amplitude into a gauge sector term ($M_{gauge}$) and a symmetry breaking term ($M_{SB}$). The first term has a "bad" high-energy behavior but it is cancelled by the second term. However, for the low energy case the second term becomes negligible if all the exchange quanta from the SB sector are heavy. Then the left over $M_{gauge}$ term are only relevant amplitudes, which are expressed as low energy theorems.

For example, in the minimal Higgs model of SM, consider the scattering $W_L^+ W_L^- \to Z_L Z_L$, for the case $s >> M_W^2$ [2,8], with the strong coupling $\lambda = \frac{M_{H^0}}{2v_0^2}$, the tree level amplitude in unitary gauge, for the scattering $W_L^+ W_L^- \to Z_L Z_L$, can be decomposed as:

$$M(W_L^+ W_L^- \to Z_L Z_L) = M_{gauge} + M_{SB}, \qquad (9)$$

where the Feynman diagrams contributing to $M_{gauge}$ and $M_{SB}$ are shown by Fig. 1 and Fig. 2 respectively. The $M_{gauge}$ amplitude is a universal function of $M_W$ and $\rho$, independent of the symmetry-breaking sector,

$$M\left(W_L^+ W_L^- \to Z_L Z_L\right) \approx \frac{g^2 s}{4M_W^2}\frac{1}{\rho} = \frac{s}{\upsilon_0^2 \rho}. \qquad (10)$$

On the other hand, the $M_{SB}$ part will have only s- channel Higgs exchange contribution in SM (Fig. 2), and therefore, for the case $s >> M_W^2$, this is given by [2,8]

$$M_{SB}\left(W_L^+ W_L^- \to Z_L Z_L\right) = -\frac{g^2 s}{4M_W^2}\frac{s}{(s - M_{H^0}^2)}. \qquad (11)$$

For the high energy case with $s >> M_{H^0}^2$ and with $\rho = 1$ at the tree level, the two contribution give,

$$M_{gauge} + M_{SB} = 0, \qquad (12)$$

thus ensuring renormalizability of the theory. On the other hand, for the low energy region with $s << M_{H^0}^2$, the symmetry breaking amplitude is negligible, i.e. $M_{SB} \approx 0$, and the scattering amplitude is dominated by $M_{gauge}$ alone.

In previous years, we have been exploring the extra $U(1)$ superstring inspired $\eta$ model emanating from $E_6$ group as a viable model for doing physics beyond SM [9-12]. This model gives reasonable negative values of S, T, U parameters [12]. It is in good agreement with SLD collaborations data on the measured of left-right symmetry parameter $A_{LR}$ [13]. The top quark mass measurements can also be accommodated in this model [9-12]. In the present paper we re-examined the low energy theorems and the unitarity bounds in the four well known extra $U(1)$ superstring inspired $E_6$ - $\chi$, $\psi$, $\eta$ and $\nu$ models[a]. The presence of additional couplings $W^{\pm} H^{\mp} Z_i$ $(i = 1,2)$ available in these models add u- channel and t-channel amplitude contributions to s- channel amplitude. We examine the impact, of these additional contributions (Fig. 3), on the low energy theorems and unitarity bounds.

[a] Here after, we will refer these models as extra U(1) models.

The plan of the paper is as follows: In Sec. II, we evaluate the scattering amplitude for the processes $W_L^+ W_L^- \to Z_{1L} Z_{1L}$, $W_L^+ W_L^- \to Z_{2L} Z_{2L}$, $W_L^+ W_L^- \to Z_{1L} Z_{2L}$, and $W_{iL}^{\pm} Z_{iL} \to W_{iL}^{\pm} Z_{iL}$ ($i$ = 1,2) in the extra $U(1)$ models. We examine the behavior of the amplitudes both at low and at high energy. At low energy, the theorems for the gauge boson scattering amplitudes remain valid but the high-energy amplitudes do not seem to satisfy the relation given Eq. (12) consistently with the requirement of present day high energy phenomenology. In Sec. III summarizing the methodology, we evaluate the coupled channels scattering matrix for two gauge bosons scattering processes involving scalar particles in the extra $U(1)$ models. In Sec. IV, we present our analysis and obtain unitarity imposed constraints on quartic scalar couplings. The bounds on low-energy parameters are obtained by evolving the relevant RGEs for scalar superpotential couplings and gauge couplings. The results of this analysis are used in Sec. V for obtaining various theoretical constraints on physical masses of top quark, lightest Higgs, D-quark and $\tan\beta$ in the extra $U(1)$ models. Finally, in Sec. VI, we mention limitations of our calculations and contrast our results with those of the corresponding ones from flipped $SU(5) \times U(1)$ given in Ref. [14].

## II. LOW – ENERGY THEOREMS IN THE EXTRA U(1) MODELS

Before we take up the actual calculations for this case, we first briefly give an outline of the extra $U(1)$ models emanating from $E_6$ group in superstring theory. Our focus is primarily on the models having gauge symmetry $SU(3)_C \times SU(2)_L \times U(1)_Y \times U(1)'$ as a subgroup of $E_6$ of rank 5. Here the $U(1)'$ represents the $U(1)_{\chi,\psi,\eta,y}$ models. It is well known [15-18] that the breakdown of $E_6$ may give rise to the effective low energy groups of rank-6 or of rank-5.

$$\begin{aligned}
E_6 &\to SO(10)\times U(1)_\chi && \text{"}rank-6\text{"}\\
&\to SU(5)\times U(1)_\chi \times U(1)_\psi && \text{"}rank-6\text{"}\\
&\to SU(3)_C\times SU(2)_L\times U(1)_Y\times U(1)_\chi\times U(1)_\psi && \text{"}rank-6\text{"}\\
&\to SU(3)_C\times SU(2)_L\times U(1)_Y\times U(1)'\ or\ U(1)_\eta && \text{"}rank-5\text{"}
\end{aligned}$$

In the group rank 5, the additional extra $U(1)$ may signal the presence of extra neutral gauge boson denoted by $Z'$ in the theory. If there is only one extra $Z$ at low energy then this $Z'$ can be parameterized as a linear combinations of $Z_\chi$ and $Z_\psi$ [16, 17], i.e.

$$Z'(Z_E) = Z_\chi \cos\theta_{E_6} + Z_\psi \sin\theta_{E_6}, \tag{13}$$

where $\theta_{E_6}$ is a free parameter having the range $-\frac{\pi}{2}\le\theta_{E_6}\le\frac{\pi}{2}$. The specific choice of $\theta_{E_6}=0$, $\frac{\pi}{2}$, $\tan^{-1}(-\sqrt{5/3})$ and $\tan^{-1}(\sqrt{15})$ corresponds to $Z_\chi$, $Z_\psi$, $Z_\eta$ and $Z_\nu$ respectively. The four representative extra $Z$ boson models are the ones most frequently analyzed in the literature, e.g., in Refs. [16,17,19] as being the canonical models which might arise in the low energy superstring motivated models. In general, as the Higgs field have non-zero $U(1)'$ charge, there would be mixing between the two neutral gauge bosons $Z$ and $Z'$. As such, the mass eigen states $Z_1$, $Z_2$ are related to $Z$ and $Z'$ through the relations

$$Z_1 = Z\cos\theta_E + Z'\sin\theta_E, \tag{14}$$

$$Z_2 = -Z\sin\theta_E + Z'\cos\theta_E, \tag{15}$$

where $\theta_E$ is the mixing angle. It is expected to be very small by the LEP and SLD Z- pole data and by other constraints [19-21].

Holdom [22] had shown that a theory which has two or more $U(1)$ gauge factors can have non-diagonal wave function in the space of $U(1)$ gauge fields. This implies that charges which are integral multiples at one scale need not be integral multiple at another scale, and their electromagnetic charges can be shifted by some amount $\varepsilon$ . Thus in a theory with two $U(1)$ factors a term in the lagrangian can appear consistently with all gauge symmetries which mixes two $U(1)$'s [18,23,24]. The pure gauge interaction Lagrangian for any arbitrary two $U(1)_a$ and $U(1)_b$ group thus becomes,

$$£ = -\frac{1}{4}F_a^{\mu\nu}F_{a\mu\nu} - \frac{1}{4}F_b^{\mu\nu}F_{b\mu\nu} - \frac{\sin\chi}{2}F_a^{\mu\nu}F_{b\mu\nu}, \tag{16}$$

which is common to all abelian and non-abelian gauge extensions. For non-abelian case, the kinetic mixing term $\chi = 0$, since $F_{b\mu\nu}$ is not gauge invariant. The kinetic mixing term with mixing determined by $\chi$ appears at some level in almost all GUT or string models due to the incomplete GUT matter representation at low energy scale, i.e., when $\sum_{chiral\,fields} Q_a^i Q_b^i \neq 0$, then a non-zero $\chi$ will be generated at one loop level [18,23]. Therefore under such circumstances the relations for $Z_{1,2}$ mass eigen states get modified as [18,23] under:

$$Z = \left(\cos\theta_E + \sin\theta_E \sin^2\theta_W \tan\chi\right)Z_1 - \left(\sin\theta_E - \cos\theta_E \sin^2\theta_W \tan\chi\right)Z_2, \tag{17}$$

$$Z' = \left(\sin\theta_E Z_1 + \cos\theta_E Z_2\right)/\cos\chi \ . \tag{18}$$

If the new $U(1)'$ results from Wilson-loop breaking mechanism directly to the rank 5 subgroup in string context, then new $U(1)'$ is indeed $U(1)_\eta$ model, which has been of our primary interest [9-12]. Our main emphasis has been to establish its viability as one of the suitable models for doing physics beyond SM, when such a need arises. However, as stated earlier, that all the extra $U(1)$ models are mutually related through the mixing angle $\theta_{E_6}$. As such our interest encompasses all the four models.

Apart from having an extra $Z'$, these models have the following Higgs structure for each generations of fermions in analogy with that for the $\eta$ model [9,25][b]:

$$H = \begin{pmatrix} H^+ \\ H^0 \end{pmatrix} = \frac{1}{\sqrt{2}} \begin{pmatrix} \phi_1 + i\phi_2 \\ \phi_3 + i\phi_4 \end{pmatrix}, \qquad \overline{H} = \begin{pmatrix} H^0 \\ H^- \end{pmatrix} = \frac{1}{\sqrt{2}} \begin{pmatrix} \overline{\phi}_1 + i\overline{\phi}_2 \\ \overline{\phi}_3 + i\overline{\phi}_4 \end{pmatrix}$$

and $$N = \frac{1}{\sqrt{2}}(\phi_{1s} + \phi_{2s}) \qquad (19)$$

The VEVs are defined as [9]

$$\langle 0|H|0\rangle = v\ , \quad \langle 0|\overline{H}|0\rangle = \overline{v}\ , \quad \langle 0|N|0\rangle = x\ . \qquad (20)$$

The gauge symmetry breaking is accomplished by the evolution of renormalization group of masses [15,25,26]. It is assumed that the third generation of fermions acquires masses through spontaneous symmetry breaking [27,28]. After symmetry breaking, one is left with three neutral Higgs, two oppositely charged Higgs and a neutral pseudoscalar Higgs. The mass eigen states of neutral Higgs are denoted by $H_1^0$, $H_2^0$, $H_3^0$ and those of charged Higgs by $H^-$, $H^+$. $\rho^0$ denotes the mass eigen state of pseudoscalar Higgs.

However, we wish to point out that in calculating the unitarity constraints using low energy theorems for the processes, namely, $W_L^+W_L^- \to Z_{1L}Z_{1L}$, $W_L^+W_L^- \to Z_{2L}Z_{2L}$, $W_L^+W_L^- \to Z_{1L}Z_{2L}$ and $W_L^\pm Z_{iL} \to W_L^\pm Z_{iL}$ ($i$ =1,2), the various couplings reported in [9,29] get slightly modified with the inclusion of gauge kinetic mixing term etc. These are given in the

---

[b] It may be emphasized that when the compactifications of the $E_8 \times E_8'$ group of superstring take place on a Calabi-Yau manifold then the resulting group can be broken directly to the group of rank 5 without having an intervening group of rank 6. As such there is no necessity of having two singlet Higgs fields. This aspects has been discussed at length by Ellis et al. in Ref. [25] (article 3.2, pp.28-32). We have made use of the same in writing Eq. (19).

Appendix. We have used these modified couplings to calculate amplitudes for various processes as shown below.

## A. The process $W_L^- W_L^+ \to Z_{1L} Z_{1L}$

The Feynman diagrams contributing to $M_{gauge}$ part of this process are similar to those given in Fig. 1 and thus giving for the $M_{gauge}$ amplitude an expression similar to Eq. (10),

$$M_{gauge}(W_L^- W_L^+ \to Z_{1L} Z_{1L}) = \frac{g^2 s}{4M_W^2}\frac{1}{\rho'}. \tag{21}$$

Here $\rho' = 1 + \delta_{\rho M}$, with the extra contributions [30] $\delta_{\rho M} = \left(\frac{M_{Z_2}^2}{M_{Z_1}^2} - 1\right)\sin\theta_E$. It is the modification that comes in extra $U(1)$ models. The value of $\delta_{\rho M}$ is expected to be negligible due to stringent constrains on the value of $\theta_E$ [19-21].

For the symmetry breaking part $M_{SB}$, apart from the s- channel diagram there are two additional diagrams (similar to Fig. 2 with $H_{SM}^0$ replaced by $H_3^0$ and modified couplings) in the t- and u- channels. These occur because of the additional presence of $W^{\pm} H^{\mp} Z_i$ vertices exclusive to those models. The expressions for various amplitudes are given below with the generalized coefficients $A_1^2$ and $B_1^2$ for $E_6$ extra $U(1)$ models:

$$iM_{SB}^s\left(W_L^- W_L^+ \to Z_{1L} Z_{1L}\right) = -\frac{g^2 s}{4M_W^2}\frac{s}{\left(s - M_{H_3^0}^2\right)} A_1^2, \tag{22}$$

$$iM_{SB}^t\left(W_L^- W_L^+ \to Z_{1L} Z_{1L}\right) = -\frac{g^2 s}{4M_W^2}\frac{t}{\left(t - M_{H^-}^2\right)} B_1^2, \tag{23}$$

$$iM_{SB}^u\left(W_L^- W_L^+ \to Z_{1L} Z_{1L}\right) = -\frac{g^2 s}{4M_W^2}\frac{u}{\left(u - M_{H^-}^2\right)} B_1^2, \tag{24}$$

with

$$A_1^2 = \left[ \frac{1}{2}\cos\beta \left\{ \cos^2\theta_E + \sin^2\theta_E x_W \tan^2\chi + \sin 2\theta_E \sqrt{x_W}\tan\chi + \frac{\sqrt{x_W}}{3} C_1 \sin 2\theta_E \sec\chi + \right.\right.$$

$$\left. \frac{2}{3}\sqrt{x_W} C_1 \sin^2\theta_E \sqrt{x_W}\tan\chi\sec\chi + \frac{1}{9} x_W C_1^2 \sin^2\theta_E \sec^2\chi \right\} U_{13} +$$

$$\frac{1}{2}\sin\beta \left\{ \cos^2\theta_E + \sin^2\theta_E x_W \tan^2\chi + \sin 2\theta_E \sqrt{x_W}\tan\chi - \frac{4}{3}\sqrt{x_W}\, C_1 \sin 2\theta_E - \right. \qquad (25a)$$

$$\left. \frac{8}{3} C_1 \sin^2\theta_E x_W \tan\chi\sec\chi + \frac{16}{9} x_W C_1^2 \sin^2\theta_E \sec^2\chi \right\} U_{23} +$$

$$\left. \frac{25}{18}\frac{x}{\upsilon} x_W \sin^2\theta_E \sec^2\chi\, U_{33} \right] \left( \cos\beta\, U_{13} + \sin\beta\, U_{23} \right),$$

$$B_1^2 = \frac{25}{36} C_1^2 x_W \sin^2 2\beta\, \frac{\sin^2\theta_E}{\cos^2\chi} \left( \frac{M_{H^-}^2}{M_{H_3^0}^2} \right), \qquad (25b)$$

where

$$\tan\beta = \frac{\bar{\upsilon}}{\upsilon},\; x_W = \sin^2\theta_W,\; C_1 = \sin\theta_{E_6} + \cos\theta_{E_6} \qquad (25c)$$

$$U_{13} = \left( \frac{36}{25}\frac{\lambda_H^2}{g^2\tan^2\theta_W} - \frac{1}{5} \right)\left( \frac{\upsilon}{x} \right), \qquad (25d)$$

$$U_{23} = \left( \frac{36}{25}\frac{\lambda_H^2}{g^2\tan^2\theta_W} - \frac{4}{5} \right)\left( \frac{\bar{\upsilon}}{x} \right), \qquad (25e)$$

$$U_{33} = 1. \qquad (25f)$$

For the low energy case with $s << M_{H_3^0}^2$ and $u, t << M_{H^\pm}^2$

$$M_{SB}^s + M_{SB}^t + M_{SB}^u \approx 0, \qquad (26)$$

and we are left with only

$$M_{gauge}\left( W_L^- W_L^+ \to Z_{1L} Z_{1L} \right) = \frac{g^2 s}{4M_W^2}\frac{1}{\rho'} = \frac{s}{\upsilon_0^2\rho'} \cong \frac{s}{\upsilon_0^2}, \qquad (27)$$

with $\upsilon_0^2 = \upsilon^2 + \bar{\upsilon}^2$ and $\rho' \approx 1$, which gives the low energy theorems for this amplitude. On the other hand, for high energy case with $s >> M^2_{H_3^0}$ and $u,t >> M^2_{H^\pm}$, the sum of the amplitudes, $M_{gauge} + M^s_{SB} + M^t_{SB} + M^u_{SB}$ can vanish as required by unitarity, if the following conditions is satisfied:

$$A_1^2 + 2B_1^2 = 1. \tag{28}$$

Since $A_1^2$ and $B_1^2$ involve many parameters, such as angles $\theta_E$, $\beta$, $\chi$ and $\theta_{E_6}$, vacuum expectation values $x, \upsilon, \bar{\upsilon}$ and masses of Higgs bosons, it is not obvious to derive a transparent constraint from Eq. (28). However, for negligible value of $\theta_E \approx 0$, $x >> \upsilon, \bar{\upsilon}$ and $M_{H^\pm} << M_{H_3^0}$ (neglecting gauge kinetic mixing effect) one gets the constraint

$$\lambda_H > \frac{5}{6} g_L \left( \frac{\sqrt{2}x}{\upsilon_0} \right)^{1/2} \tan\theta_W . \tag{29}$$

In order to keep $\lambda_H$ in the perturbative region [15,29] i.e. $\frac{\lambda_H^2}{4\pi} < 1$, the above constraint requires, $x < 24682\ GeV$ for $\sin^2\theta_W \cong 0.231$ and remains unaltered for different $E_6$ extra $U(1)$ models. This requirement can be satisfied by all reasonable GUTs. As no lower limit on $x$ seems to exist there is no possibility to estimate $\lambda_{H(\min)}$ from Eqs. (29). Therefore, low energy theorems are seen to be satisfied both for low as well as for the high energy case for this process.

## B. The process $W_L^- W_L^+ \to Z_{2L} Z_{2L}$

Proceedings in a manner similar to that used for the process $W_L^- W_L^+ \to Z_{1L} Z_{1L}$, we obtain expressions for different amplitudes. The expressions are similar to those obtained for the process $W_L^- W_L^+ \to Z_{1L} Z_{1L}$ except for slight changes in the couplings and coefficients $A_2^2$ and $B_2^2$ for the extra $U(1)$ models:

$$M_{gauge}\left(W_L^- W_L^+ \to Z_{2L}Z_{2L}\right) = -\frac{g^2 s}{4M_W^2}\frac{1}{\rho'}, \tag{30a}$$

$$M_{SB}^s\left(W_L^- W_L^+ \to Z_{2L}Z_{2L}\right) = -\frac{g^2 s}{4M_W^2}\frac{s}{\left(s - M_{H_3^0}^2\right)}\left(\frac{M_{Z_1}^2}{M_{Z_2}^2}\right)A_2^2, \tag{30b}$$

$$M_{SB}^t\left(W_L^- W_L^+ \to Z_{2L}Z_{2L}\right) = -\frac{g^2 s}{4M_W^2}\frac{t}{\left(t - M_{H^-}^2\right)}\left(\frac{M_{Z_1}^2}{M_{Z_2}^2}\right)B_2^2, \tag{30c}$$

$$M_{SB}^u\left(W_L^- W_L^+ \to Z_{1L}Z_{1L}\right) = -\frac{g^2 s}{4M_W^2}\frac{u}{\left(u - M_{H^-}^2\right)}\left(\frac{M_{Z_1}^2}{M_{Z_2}^2}\right)B_2^2, \tag{30d}$$

with

$$\begin{aligned}A_2^2 = C_1^2\Bigg[&\frac{1}{2}\cos\beta\Big\{\sin^2\theta_E + \cos^2\theta_E x_W\tan^2\chi - \sin 2\theta_E\sqrt{x_W}\tan\chi - \frac{1}{3}\sqrt{x_W}\sin 2\theta_E\sec\chi + \\ &\quad \frac{2}{3}\cos^2\theta_E x_W\tan\chi\sec\chi + \frac{1}{9}x_W\sin^2\theta_E\sec^2\chi\Big\}U_{13} + \\ &\frac{1}{2}\sin\beta\Big\{\sin^2\theta_E + \cos^2\theta_E x_W\tan^2\chi - \sin 2\theta_E\sqrt{x_W}\tan\chi + \frac{8}{3}\sqrt{x_W}\sin 2\theta_E\sec\chi - \\ &\quad \frac{8}{3}\cos^2\theta_E x_W\tan\chi\sec\chi + \frac{16}{9}x_W\cos^2\theta_E\sec^2\chi\Big\}U_{23} + \\ &\frac{25}{18}\frac{x}{v}x_W\cos^2\theta_E\sec^2\chi\, U_{33}\Bigg]\left(\cos\beta\, U_{13} + \sin\beta\, U_{23}\right),\end{aligned} \tag{30e}$$

$$B_2^2 = \frac{25}{36}C_1^2 x_W\sin^2 2\beta\,\frac{\cos^2\theta_E}{\cos^2\chi}\left(\frac{M_{H^-}^2}{M_{H_3^0}^2}\right). \tag{30f}$$

For the low energy case, with $s << M_{H_3^0}^2$ and $u,t << M_{H^\pm}^2$, we find,

$$M_{SB}^s + M_{SB}^t + M_{SB}^u \approx 0,$$

and with $\rho' \approx 1$

$$M_{gauge} = \frac{gs}{4M_W^2} = \frac{s}{v_0^2},$$

implying the validity of the low energy theorems [2,3]. On the other hand, for the high energy case, with $s >> M^2_{H_3^0}$ and $u,t >> M^2_{H^\pm}$, the sum of the amplitudes, $M_{gauge} + M^s_{SB} + M^t_{SB} + M^u_{SB}$ vanish (as required by unitarity), if the following condition is satisfied,

$$A_2^2 + 2B_2^2 = \frac{M^2_{Z_2}}{M^2_{Z_1}} . \tag{31}$$

Again $A_2^2$, $B_2^2$ involve too many parameters and as such Eq. (31) does not lead to useful constraints. For the special case of $\theta_E \approx 0$, $x >> \upsilon, \bar{\upsilon}$, $M_{H^\pm} << M_{H_3^0}$ and neglecting gauge kinetic mixing effect, it gives constraint on $x$ in the form of relation,

$$x \le \left| \upsilon_0 \sqrt{\left(11.287C_1^2\lambda_H^2 - 0.199C_1^2 - 4.193\right)} \right|, \tag{32}$$

The upper bounds on $x$ in are given in Table I and do not differ significantly amongst the four extra $U(1)$ models. This may or may not lead to acceptable $Z_2$ mass that may satisfy the phenomenological constraints arising out of flavor changing neutral currents (FCNC) due to presence of other parameters like $\upsilon$, $\bar{\upsilon}$.

Likewise for the processes $W_L^- W_L^+ \to Z_{iL} Z_{iL}$, $W_L^\pm Z_{iL} \to W_L^\pm Z_{iL}$ (i = 1, 2), we have verified that for the high energy case we do not arrive at a transparent picture regarding the unitarity bound as in the above case of $W_L^- W_L^+ \to Z_{2L} Z_{2L}$. As such we lare led to believe that the low energy framework [1,10] for investigating unitarity. Bu the latter method in extra U(1) models will involve calculations of amplitudes of very large number of processes including supersymmetric ones, and the framework will become almost unwieldy to arrive at a sensible unitarity constraint limit. We have therefore attempted to make use of an elegant and relatively much less cumbersome method due to Durand & Lopez [14] which they have used in the case of flipped *SU(5)×U(1)* superstring GUT model[c]. This method is suitable for the case of supersymmetric GUT model for testing unitarity and to extract limits on superpotential couplings there from. In the next section, we have made use of method [14] in the case of extra $U(1)$ models under discussion.

## III. SUPERPOTENTIAL, QUARTIC SCALAR COUPLINGS AND SCATTERING AMPLITUDES

The method of Ref. [14] consists of (i) setting all parameters to zero that cannot contribute to the high energy limit of the tree-level scattering amplitudes, and thereby can not contribute to potential violations of unitarity, thus setting to zero all soft supersymmetry breaking terms, (ii) calculate only quartic scalar interactions using the Goldstone boson equivalence theorem [2, 31] thus neglecting all gauge couplings. (iii)As fermionic interactions in supersymmetric theories lead as expected to the same unitarity bounds as the scalar interactions [32] and need not be calculated separately, (iv) obtain the high energy unitarity constraints and regard them holding at the Grand unified scale $M_G$, and finally, (v) evolve the constraints at low energy using the appropriate renormalization group equations, thus obtaining in general much stronger constraints on the parameters of the low energy models. In the following, we use the aforesaid method for ascertaining unitarity bounds in specific $E_6$ extra $U(1)$ models under consideration.

We make use of the following $E_6$ superpotential (common for$\chi, \psi, \eta$ and $\nu$ models)for energies between $M_G$ and $M_Z$ [15,25,26]

$$W = \lambda_t t^c qH + \lambda_D ND^c D + \lambda_b b^c q\overline{H} + \lambda_H N\overline{H}H \tag{33}$$

[c] The use of this method for non-supersymmetric GUT models is not ruled out [14].

where $\lambda's$ denote various couplings, and $q = \begin{pmatrix} t \\ b \end{pmatrix}$ for the third generation of quarks. Using the procedure of Ref. [14], we obtain the following expression for quartic potential, $V_{quartic}$ :

$$\begin{aligned} V_{quartic} = & \left(\lambda_b b\phi_1^0\right)^2 + \left(\lambda_b b^c\phi_1^0 - \lambda_t t\phi_2^+\right)^2 + \left(\lambda_t t\phi_2^0\right)^2 + \left(\lambda_b b^c\phi_1^- - \lambda_t t^c\phi_2^0\right)^2 + \\ & \left|\lambda_t b^c b + \lambda_H \phi_2^0\phi_3\right|^2 + \left|\lambda_t t^c b + \lambda_H \phi_1^0\phi_3\right|^2 + \left|\lambda_t b^c t + \lambda_H \phi_2^+\phi_3\right|^2 + \\ & \left|\lambda_t t^c b + \lambda_H \phi_1^-\phi_3\right|^2 + \left|\lambda_H\left(\phi_1^0\phi_2^0 - \phi_1^-\phi_2^+\right) + \lambda_D D^c D\right|^2 + \left|\lambda_D \phi_3 D^c\right|^2 + \\ & \left|\lambda_D \phi_3 D\right|^2 \end{aligned} \qquad (34)$$

In order to rewrite $V_{quartic}$ in terms of a set of charged and neutral fields, which are the physical fields in the absence of mixing, i.e. at high energy, we re-define the fields by making use of equivalence theorem [31] as follows

$$H_1 = \mathrm{Re}\left(\phi_1^0 + \phi_2^0\right), \quad H_2 = \mathrm{Im}\left(\phi_1^0 + \phi_2^0\right), \quad H_3 = \sqrt{2}\,\mathrm{Re}\phi_3,$$

$$H_0 = \mathrm{Re}\left(\phi_2^0 - \phi_1^0\right), \quad H^\pm = \frac{\left(\phi_1^\pm + \phi_2^\pm\right)}{\sqrt{2}}, \qquad (35)$$

with the Goldstone bosons,

$$Z = \mathrm{Im}\left(\phi_2^0 - \phi_1^0\right), \quad Z' = \sqrt{2}\,\mathrm{Im}\phi_3,$$

$$W^- = \frac{\left(\phi_2^- - \phi_1^-\right)}{\sqrt{2}}, \quad W^+ = \frac{\left(\phi_2^+ - \phi_1^+\right)}{\sqrt{2}}. \qquad (36)$$

This gives

$$\begin{aligned} V_{quartic} = & V_b(\lambda_b) + V_D(\lambda_D) + V_t(\lambda_t) + V_H(\lambda_H) + V_{b,H}(\lambda_b, \lambda_H) + \\ & V_{t,H}(\lambda_t, \lambda_H) + V_{D,H}(\lambda_D, \lambda_H), \end{aligned} \qquad (37)$$

with

$$V_b(\lambda_b) = \frac{\lambda_b^2}{4}\Big[b_+ b_- \left(H_0^2 + H_1^2 + H_2^2 + Z^2 - 2H_1H_0 - 2H_2Z\right) +$$
$$b_+^c b_-^c \Big(H_0^2 + H_1^2 + H_2^2 + Z^2 - 2H_1H_0 - 2H_2Z + 2H^+H^- +$$
$$2W^+W^- - 2H^+W^- - 2H^-W^+\Big) + 4b_+^c b_- \left(b_+^c b_- + t_-^c t_+\right)\Big], \qquad (38a)$$

$$V_t(\lambda_t) = \frac{\lambda_t^2}{4}\Big[t_+ t_- \left(H_0^2 + H_1^2 + H_2^2 + Z^2 + 2H_1H_0 + 2H_2Z\right) +$$
$$t_+^c t_-^c \Big(H_0^2 + H_1^2 + H_2^2 + Z^2 + 2H_1H_0 + 2H_2Z + 2H^+H^- +$$
$$2W^+W^- + 2H^+W^- + 2H^-W^+\Big) + 4t_+^c t_- \left(t_+^c t_- + b_+^c b_-\right)\Big], \qquad (38b)$$

$$V_D(\lambda_D) = \lambda_D^2 D_+^c D_+ D_-^c D_- , \qquad (38c)$$

$$V_H(\lambda_H) = \frac{\lambda_H^2}{16}\Big[\Big(6H_1^2H_3^2 + 6H_2^2H_3^2 + 6H_0^2H_3^2 + 6Z^2H_3^3 + 4H_1H_0H_3^2 +$$
$$4H_2H_3^2Z + 8H_+H^-H_3^2 + 8W^+W^-H_3^2\Big) - \Big(6H_1^2Z'^2 +$$
$$6H_2^2Z'^2 + 6H_0^2Z'^2 + 6Z^2Z'^2 + 4H_1H_0Z'^2 + 4H_2ZZ'^2 +$$
$$8H^+H^-Z'^2 + 8W^+W^-Z'^2\Big) + \Big(H_0^2H_0^2 + H_1^2H_1^2H_2^2H_2^2 +$$
$$Z^2Z^2 + 2H_1^2H_2^2 + 2H_1^2Z^2 + 2H_0^2H_2^2 + 2H_0^2Z^2 -$$
$$2H_0H_1H_0H_1 - 2H_2ZH_2Z - 8H_1H_0H_2Z\Big) +$$
$$\Big(4H^+H^-H^+H^- + 4W^+W^-W^+W^- - 4H^-W^+H^-W^+ -$$
$$4H^+W^-H^+W^-\Big) + \Big(-4H_1^2H^+H^- - 4H_1^2W^-W^+$$
$$+ 4H_1^2H^-W^+ + 4H_1^2H^+W^- + 4H_2^2H^+H^- + 4H_2^2W^-W^+ -$$
$$4H_2^2H^-W^+ - 4H_2^2H^+W^- + 4H_0^2H^+H^- + 4H_0^2W^-W^+ -$$
$$4H_0^2H^-W^+ - 4H_0^2H^+W^- + 4Z^2H^+H^- + 4Z^2W^-W^+ -$$
$$4Z^2H^-W^+ - 4Z^2H^+W^- + 8H_0ZH^+H^- + 8H_0ZW^-W^+ -$$
$$8H_0ZH^-W^+ - 8H_0ZH^+W^-\Big) + \Big(8iH_1H_2 - H^+H^- - W^-W^+$$
$$+ H^-W^+ + H^+W^-\Big)\Big], \qquad (38d)$$

$$V_{b,H}(\lambda_b, \lambda_H) = (\lambda_b, \lambda_H)\Bigg[\frac{b_+^c b_-}{\sqrt{2}}\big[H_0H_3 + H_1H_3 - H_2Z' - ZZ' +$$
$$i\left(H_2H_3 + ZH_3 + H_1Z' + H_0Z'\right)\big] +$$
$$b_c^+ t_+ \left[H^+H_3 + W^+H_3 + i\left(H^+Z' + W^+Z'\right)\right] - \Bigg], \qquad (38e)$$

$$V_{t,H}(\lambda_t, \lambda_H) = (\lambda_t, \lambda_H)\Bigg[\frac{t_-^c t_+}{\sqrt{2}}\big[-H_0H_3 + H_1H_3 - H_2Z' + ZZ' +$$
$$i\left(H_2H_3 - ZH_3 + H_1Z' - H_0Z'\right)\big] +$$
$$t_-^c b_- \left[H^-H_3 - W^-H_3 + i\left(H^-Z' - W^-Z'\right)\right] - \Bigg], \qquad (38f)$$

$$V_{D,H}(\lambda_D,\lambda_H) = \frac{(\lambda_D,\lambda_H)}{2}\left[D^c_- D_+\left[- H_0^2 + H_1^2 - H_2^2 + Z^2 - 2H^+H^- + 2W^+W^- - 2H^-W^+ - 2H^+W^- + 2i(H_0H_1 - H_0Z)\right]\right], \tag{38g}$$

In order to calculate the coupled channels scattering matrix for two body scattering process involving scalar particles in the extra $U(1)$ models, we follow the calculational steps of Durand and Lopez [14] outlined in the beginning of this section III. We use only the $V_{quartic}$ parts given by Eqs. (38a)-(38g). Furthermore, we use properly normalized *j = 0*, partial wave amplitudes. The scattering matrix for the coupled channel breaks into four such matrices labeled $T_1$, $T_2$, $T_3$, $T_4$ with the following channels:

$$T_1: \quad W^-W^+, ZZ, Z'Z', H^-H^+, H_0H_0, H_1H_1, H_2H_2, H_3H_3, W^-H^+, W^+H^-, H_1H_0, H_2Z, b^c_+b^c_-, b_+b_-, t^c_+t^c_-, t_+t_-, D^c_-D_+, H_0Z, H_1H_2 \tag{39a}$$

$$T_2: \quad b^c_+b_-, t^c_-t_+, H_0H_3, H_1H_3, H_1Z', H_2H_3, H_2Z', H_3Z, ZZ' \tag{39b}$$

$$T_3: \quad t^c_-b_-, H^-H_3, W^-H_3, H^-Z', W^-Z' \tag{39c}$$

$$T_4: \quad b^c_+t_+, H^+H_3, W^+H_3, H^+Z, W^+Z' \tag{39d}$$

These matrices are shown respectively in Table **II, III, IV** and **V** with the entries in the order of the channel label given in Eqs. (39a)-39(d).

## IV. ANALYSIS AND UNITARITY CONSTRAINTS

We now impose the unitarity constraint [14] that the largest eigenvalue of the coupled channel scattering matrix must be less than unity in magnitude. This determines a four dimensional parametric space of the parameters $\lambda_t$, $\lambda_b$, $\lambda_H$, $\lambda_D$ and leads to the maximum unitarity constraints on the following quartic scalar superpotential couplings at GUT scale (Fig. 4):

$$\lambda_{t(\max.)}(M_G) = 5.574; \qquad \lambda_{b(\max.)}(M_G) = 5.574 \tag{40a}$$

$$\lambda_{H(\max.)}(M_G) = 3.011; \qquad \lambda_{D(\max.)}(M_G) = 3.545. \tag{40b}$$

These bounds remain unchanged in the case of four extra $U(1)$ models.

We have shown in Fig. 4 the projection of allowed region onto the subspace of scalar superpotential couplings $\lambda_t$, $\lambda_H$, $\lambda_D$ for $\lambda_b = 0$ for the energy range $M_Z < \sqrt{s} < M_G$. The thin lines sketched in the Fig. 4 are on the unitarity surface, which has the largest eigen value equals to unity. The parameters $\lambda_t$, $\lambda_D$ are not coupled in the quartic potential, Eqs. (38) and therefore the unitarity boundary plane is a rectangle for $\lambda_b = \lambda_H = 0$. The $\lambda_b, \lambda_t$ have identical values since the superpotential given by Eq. (33) is symmetric under the transformation $\lambda_t \leftrightarrow \lambda_b$, $b \leftrightarrow t$, and $H \leftrightarrow \overline{H}$. It is interesting to note that for $\lambda_b \neq 0$ ( $\lambda_t = 0$), the unitarity surface remains unaltered and will appear identical to that shown in Fig. 4. Further, these bounds enable us to estimate the upper bounds on the Yukawa couplings at weak scale ($M_Z$) in each $E_6$ models (neglecting kinetic mixing) with the flow of RGE.

The bounds on $\lambda_t$, $\lambda_H$ at weak scale are of special interest since they lead to the upper bounds on top quark mass and lightest neutral scalar Higgs mass (as discussed in Section V). Therefore, to obtain the bounds on the low energy parameters, we evolve the relevant RGE for scalar superpotential couplings ($\lambda$'s) and gauge couplings from GUT scale ($M_G$) to weak scale ($M_Z$). This is consistent with our motivation of exploring the physics of extra $U(1)$ models at the scale $M_Z$.

The one loop RGE for the gauge couplings of the $SU(3)_C \times SU(2)_L \times U(1)_Y \times U(1)'$ superstring inspired $E_6$ extra $U(1)$ model are [15,16,25,26] $\left(\alpha_a = g_a^2/4\pi\,;\ a = 1, 2, 3, E\right)$,

$$\frac{d\alpha_a}{dt} = b_a \alpha_a^2, \Rightarrow \quad \frac{dg_a}{dt} = \frac{b_a}{8\pi} g_a^3, \tag{41}$$

where $t = \frac{1}{2\pi}\ln\left(\frac{\sqrt{s}}{M_Z}\right)$. The analytical solutions for the gauge couplings are

$$g_a^2(t) = \left[\frac{g_a^2(0)}{1 - \frac{b_a}{4\pi} g_a^2(0)t}\right], \tag{42}$$

where $t = t_G = 5.256$ at $\sqrt{s} = M_G$. At unification scale, one should expect the equality of the gauge couplings (universality of g's) i.e.

$$g_1(t_G) = g_z(t_G) = g_3(t_G) = g_E(t_G) = 1.218, \tag{43}$$

where $M_G \approx 2\times 10^{16}\ GeV$.

The RGE of superpotential couplings $\lambda_t, \lambda_b, \lambda_H, \lambda_D$ evolve according to the following equations [11,15,25,26]

$$\frac{d\lambda_t}{dt} = \frac{\lambda_t}{8\pi}\left[6\lambda_t^2 + \lambda_b^2 + \lambda_H^2 - \frac{16}{3}g_3^2 - 3g_2^2 - \frac{13}{15}g_1^2 - 2g_E^2\left\{Q_1'^2 + Q_Q'^2 + Q_u'^2\right\}\right], \tag{44a}$$

$$\frac{d\lambda_b}{dt} = \frac{\lambda_b}{8\pi}\left[\lambda_t^2 + 6\lambda_b^2 + \lambda_H^2 - \frac{16}{3}g_3^2 - 3g_2^2 - \frac{7}{15}g_1^2 - 2g_E^2\left\{Q_2'^2 + Q_Q'^2 + Q_d'^2\right\}\right], \tag{44b}$$

$$\frac{d\lambda_H}{dt} = \frac{\lambda_H}{8\pi}\left[3\lambda_t^2 + 3\lambda_b^2 + 4\lambda_H^2 + 3\lambda_D^2 - 3g_2^2 - \frac{3}{5}g_1^2 - 2g_E^2\left\{Q_1'^2 + Q_2'^2 + Q_N'^2\right\}\right], \tag{44c}$$

$$\frac{d\lambda_D}{dt} = \frac{\lambda_D}{8\pi}\left[2\lambda_H^2 + 5\lambda_D^2 - \frac{16}{3}g_3^2 - \frac{4}{15}g_1^2 - 2g_E^2\left\{Q_N'^2 + Q_D'^2 + Q_{D^c}'^2\right\}\right]. \tag{44d}$$

The renormalized quantum numbers of chiral supermultiplets corresponding to extra $U(1)$ groups are listed in Table VI. The one-loop beta functions involved in RGEs of gauge couplings for $E_6 - \chi$, $\psi$, $\eta$ and $\nu$ models are used from reference [17]. This enables us to obtained values of the gauge couplings at weak scale. These are given in Table VII.

In order to obtain the unitarity constraints on the Yukawa couplings at weak scale, we evolved the region given in Fig. 4 from GUT scale to weak scale with the use of RGEs of Yukawa couplings and gauge couplings. The obtained bounds on Yukawa couplings at weak scale are listed in Table VII for extra $U(1)$ models. It is interesting to notice that the unitarity bounds on $\lambda_{t,b,H,D}$ are in the good vicinity of each other in the ***27*** representation of $\chi$ , $\psi$ , $\eta$ and $\nu$ models and are different for $\chi(16)$ due to the appreciable difference in the values of its beta functions with respect to those of others.

For the particular case of $E_6$-$\eta$ model ($\beta_E = 9.64$), the unitarity surface of Yukawa couplings ($\lambda_{t,H,D}$) at weak scale as shown in Fig. 5. The flows of Yukawa couplings is obtained for the various sets of Yukawa couplings from the unitarity surface at GUT scale. We have checked that the behavior of the Yukawa couplings at weak scale for the $\chi$ , $\psi$ , and $\nu$ models are found almost identical with those of $\eta$ model. This is reflected from Table VII. At weak scale, the bound on a particular coupling is also obtained by setting other Yukawa couplings ($\lambda$'s) to zero in the corresponding RGEs given in Eqs. (44) [11], where we utilize the concept of critical couplings.

For the non-zero value of $\lambda_b$ ($\lambda_t = 0$), the bounds on $\lambda_{b(\max.)}$ are very close to the bounds on $\lambda_{t(\max.)}$ at weak scale. It follows from the fact that the renormalization group Equations (44a), (44b) differ only in $Q'$ contribution which is almost negligible and the values of $\lambda_{t(\max.)}$ and $\lambda_{b(\max.)}$ at $M_G$ are equal.

The projections of unitarity surface of $\lambda_t$, $\lambda_H$ at $M_G$ and $M_Z$ are shown in Fig. 6(a) and Figs. 6(b)-6(f) for the $E_6$- $\chi(16)$, $\chi(27)$, $\psi$ , $\eta$ and $\nu$ models respectively. For $\lambda_b(M_G) = 0, \lambda_D(M_G) = 0$, the unitary bounded region is shown by the curve (i) at both $M_G$ and $M_Z$ scale, which is further reduced for the non-zero value of $\lambda_b$ at GUT, i.e. when $\lambda_b(M_G) = \lambda_t(M_G)$, and thus the effect of the evolutions in the $\lambda_t$, $\lambda_H$ plane are shown by the curve (ii). These two curves are connected by the evolution of RGEs from GUT scale to weak scale. The obtained bounds on $\lambda_t$ and $\lambda_H$ in $E_6$ models under these conditions are parameterized by $\lambda_t'$ and $\lambda_H'$ in Table VIII.

Apart from working out the flows of various couplings [14], it is interesting to solve Eqs. (44) analytically for a particular $\lambda_i (i = t, b, H, D)$ with all the other $\lambda$'s being put to zero. One finds a singular point in the solution beyond which the corresponding coupling diverges at $M_G$. The Eqs. (44) may be written as:

$$\frac{d\lambda_i}{dt} = \frac{\lambda_i}{8\pi}\left[\kappa_{ii}\lambda_i^2(t) - \sum_{a=1}^{E} C_{ia} g_a^2(t)\right], \tag{45}$$

and will have the solution [11, 14]

$$\lambda_i^2(t) = \frac{\lambda_i^2(0)}{\left(1 - \frac{\kappa_{ii}}{4\pi}\lambda_i^2(0) G_i(t)\right)} \prod_{a=1}^{E} \left[1 - \frac{b_a}{4\pi} g_a^2(0) t\right]^{\frac{C_{ia}}{b_a}}, \tag{46}$$

with
$$G_i(t) = \int_0^t \prod_{a=1}^{E} \left[1 - \frac{b_a}{4\pi} g_a^2(0) t'\right]^{\frac{C_{ia}}{b_a}} dt'. \tag{47}$$

where
$$C_{ia} = \begin{bmatrix} \frac{13}{15} & 3 & \frac{16}{3} & 2\left(Q_1'^2 + Q_Q'^2 + Q_u'^2\right) \\ \frac{7}{15} & 3 & \frac{16}{3} & 2\left(Q_2'^2 + Q_Q'^2 + Q_d'^2\right) \\ \frac{3}{5} & 3 & 0 & 2\left(Q_1'^2 + Q_2'^2 + Q_N'^2\right) \\ \frac{4}{15} & 0 & \frac{16}{3} & 2\left(Q_N'^2 + Q_D'^2 + Q_{D^C}'^2\right) \end{bmatrix}$$

and
$$\kappa_{ii} = \begin{pmatrix} 6 & 1 & 1 & 0 \\ 1 & 6 & 1 & 0 \\ 3 & 3 & 4 & 3 \\ 0 & 0 & 2 & 5 \end{pmatrix} \text{ for } a = 1,2,3E \text{ and } i = t, b, H, D.$$

Here $\lambda_i(0)$ denotes the values $\lambda_i$ at $t = 0$ i.e. at $\sqrt{s} = M_Z$ and $b_a$, $g_a^2(0)$ are already given in Table VII. The couplings $\lambda_i(t)$ diverge at $\sqrt{s} = M_G$, if

$$\lambda_{i.c} = \lambda_i(t = 0) = \left[\frac{4\pi}{\kappa_{ii} G_i(t_G)}\right]^{\frac{1}{2}} . \tag{48}$$

$\lambda_{i,c}$ in this case are called critical couplings and are related to the triviality of the scalar field theories [11]. Following the procedure given in reference [11, 14], we obtain the critical bounds for $i = t, b, H, D$ for the extra $U(1)$ models and these are given in Table IX. We notice that these values are fairly close to the corresponding unitarity bounds given in Table VII at the weak scale. *This implies that one cannot violate the low energy unitarity without at the same time having the coupling diverge for* $\sqrt{s} \leq M_G$, *and thereby forbidding perturbative treatment for the grand unified model contrary to the expectation of such a treatment for the very existence of such models.*

## V. BOUNDS ON PHYSICAL MASSES

The bounds on the top quark and lightest neutral Higgs boson mass in the extra U(1) models are obtained as follows:

### A. Bounds on top quark mass

The top quark mass is given by the following relation

$$m_t = \lambda_t v = \lambda_t v_0 \sin\beta \ . \tag{49a}$$

This implies that

$$m_t \leq \lambda_{t(\max.)} v_{\max.} , \tag{49b}$$

which gives the absolute upper bounds on its mass and are given in Table X. These clearly show that in the models with ***<u>27</u>*** representations, the bounds are very close to each other than that of the ***16*** representation. Here the upper bound on $\lambda_t$ at the weak scale $(M_Z)$ is obtained from curve "(i)" in Figs. 6(b)-6(f) when all other couplings are put to zero. In each model, the upper bound on top quark mass is independent of the value of $\tan\beta = \frac{\bar{v}}{v}$ and is fairly close to the observed experimental value.

The problem of the pronounced difference between the top and bottom quark mass can be understood either by the large difference in the Yukawa couplings $\lambda_t, \lambda_b$ with $v \approx \bar{v}$ or to a similar difference between the VEVs $v, \bar{v}$ with $\lambda_t \approx \lambda_b$. The first possibility is not possible according to the theoretical prejudices in string theory. The latter possibility is more appealing and viable by requiring the $\lambda_t \approx \lambda_b$ at higher energies i.e. at GUT scale. With this choice, the value of $\lambda_{t(\max.)}$ is reduced by ~7% at weak scale, which further strengthens the bounds on top quark mass listed in Table X.

The bottom-quark mass is determined using $m_b = \lambda_b \bar{v} \leq \lambda_{b(\max.)} \bar{v}$, where $\lambda_{b(\max.)} = 1.278$ for the $\eta$ case. Thus, one can have $\bar{v}_{(\min.)} \geq \frac{m_b}{\lambda_{b(\max.)}} = 3.497\,GeV$ for $m_b = 4.5\ GeV$. This gives $v = \sqrt{v_0^2 - \bar{v}^2} \cong 173.96\ GeV$, which does not effect substantially the result for $m_t$ as given by Eq. (49). Almost identical conclusions follow for the other $E_6$ models.

### B. Bounds on $\tan\beta$

The bounds on $\tan\beta$ may also be obtained with the use of derived unitarity bounds on $\lambda_{t(\max.)}$ and $\lambda_{b(\max.)}$ at weak scale ( given in Table VII), by making use of the following relations [14]:

$$\tan\beta = \frac{\bar{v}}{v} < \left[\frac{\lambda^2_{t(\max.)} v^2_0}{m^2_{t(\min.)}} - 1\right]^{\frac{1}{2}}, \tag{50}$$

and

$$\tan\beta = \frac{\bar{v}}{v} > \left[\frac{\lambda^2_{b(\max.)} v^2_0}{m^2_b} - 1\right]^{-\frac{1}{2}}. \tag{51}$$

The upper bounds on $\tan\beta$ corresponding to the top quark masses [33-35] are listed in Table XI for the two cases: (i) $\lambda_b(M_G) = 0$ and (ii) $\lambda_t = \lambda_b\,(M_G)$. These reveal that $\tan\beta$ should not be greater than one in the case of four extra $U(1)$ superstring inspired $E_6$ models except for the case of $\chi\,(16)$. The Eq. (51) leads to the lower bounds on $\tan\beta$ in accordance with the presently known value of bottom quark mass $(m_b)$ and are given in Table XII.

### C. Bounds on lightest neutral Higgs boson mass

To estimate the upper bound on lightest neutral CP-even Higgs mass in the extra U(1) models, we make use of the following relation, due to Drees [36]

$$m^{2(tree)}_{H^0_2} \le M^2_Z \cos^2 2\beta + v^2_0\left[\lambda^2_H \sin^2 2\beta + \frac{g'^2}{18}\left(4\sin^2\beta + \cos^2\beta\right)^2\right], \tag{52}$$

which holds good in the limit $x^2 >> v^2 >> \bar{v}^2$. But Drees [37] has pointed out that the above relation gets modified due to the maximum one loop radiative contribution from top-"s-top" system to the Higgs mass through the relation [38]

$$\Delta^2_{1loop} = M^2_Z\left[1 + \frac{16}{9}\frac{g'^2}{g^2_Z}\right] + \frac{3\alpha}{2\pi\cos^2\theta_W \sin^2\theta_W}\frac{m^4_t}{M^2_Z}\ln\frac{m^2_{\tilde{t}}}{m^2_t} \tag{53}$$

where $g' = \sqrt{\tfrac{3}{5}}g_1$ and $g_Z = \sqrt{g^2_2 + g'^2}$.

Thus, finally the modified Higgs mass relation comes as

$$m^2_{H^0_2} \leq m^{2(tree)}_{H^0_2} + \Delta^2_{1loop} . \tag{54}$$

We use our results obtained so far from the unitarity constraints on $\lambda_t, \lambda_H$ to determine the $m_{H^0_2}$, where the maximum value of $\tan\beta$ is given by Eq. (50). The Eq. (54) illustrates that the bounds on lightest CP-even Higgs mass beyond SM increases due to $m_t^4$ enhancement of one-loop radiative correction considering minimal/negligible top-"s-top" mixing.

Thus the upper bounds on $m^{tree}_{H^0_2}$ (at tree level) and $m_{H^0_2}$ (inclusion of "s-top" contributions) for the recent measurements of top quark mass [33-35] are show in Table V, when both top and bottom quark couplings are completely independent of each other at GUT scale. These bounds are further reduced by less than 1% on assuming the equality of $\lambda_t$ and $\lambda_b$ at GUT.

### D. Bounds on D-quark mass

It is also possible to estimate the bounds on additional charge (-1/3) colour triplet particle *D* of spin ½, exclusively present in extra $U(1)$ models, which make use of compactifications on a Calabi-Yau manifold [39]. The mass of *D*-quark is given by

$$m_D \cong \lambda_D \langle 0|N|0 \rangle \cong \lambda_D x . \tag{55}$$

This implies that the mass of *D*-quark is completely independent of those of ordinary quark and leptons the latter being proportional not only to different Yukawa couplings but also to different VEVs. The precise bound on D-quark mass cannot be obtained because (i) there is still no precise observation of correct electroweak breaking scale for which the singlet VEV i.e. *x* is responsible and (ii) there is no upper bounds on $Z'$ mass, which is also related to the singlet VEV i.e. $x \cong \langle 0|N|0 \rangle$. Since, if upper bound on $Z'$ mass is found then the same on *x* can be obtained for the favored values of $\tan\beta$ i.e. $\leq 1$ of these models. This may then lead to a bound on the mass of *D*-quark. Latter issue seems more feasible and thus should be explored in future colliders, namely, at LHC [40] and LEP [41]. A success in this attempt could decisively be a milestone in the understanding of electroweak breaking scale as also a good deal of new physics beyond the standard model including the physics of $Z'$.

We have calculated plausible range of *D*-quark mass on varying $Z_2$ mass for the favored value of $\tan\beta$ in the extra $U(1)$ model with the use of following relations:

$$m_{D(\max.)} \cong \lambda_{D(\max.)}\, x, \tag{56}$$

where *x*, the VEV of singlet, and is given by [11, 25]

$$x = \left(\frac{\upsilon}{5}\right)\left[\left(\frac{9(1+\tan^2\beta)}{\sin^2\theta_W}\right)\left(\frac{M_{Z_2}^4 - 2M_{Z_2}^2 M_Z^2 - 4a^2 M_Z^4}{2M_Z^2 M_{Z_2}^2 - 4M_Z^4}\right) - \tan^2\beta - 16\right]^{1/2}, \tag{57}$$

with

$$a = \frac{1}{3}\sin\theta_W\left(\frac{4-\tan^2\beta}{1+\tan^2\beta}\right). \tag{58}$$

Keeping in view the possibility of the detection of $Z'$ at future Colliders [42], the possible values of the ratio $\left(\frac{x}{\upsilon}\right)$ are shown in Fig. 7(a) for various $M_{Z_2}$ masses within the limit on $\tan\beta$ in this model. This may enables us to ascertain the value of singlet VEV i.e. *x* along with the use of $\upsilon \approx \upsilon_0 \cong 174\ GeV$. With $\lambda_{D(\max.)}(M_Z) = 1.312$, Fig. 7(b) shows the variation of *D*-quark mass with respect to $M_{Z_2}$. But since no upper bounds are available on $M_{Z_2}$, it is not possible to prescribe upper bound on *D*-quark mass. However, in order to have some order of magnitude idea, and keeping in view of the present day experimental situation [42] we have taken $M_{Z_2}$ to be the order of 1 *TeV*, which leads to the *D*-quark mass to be the order of 3 *TeV* [25].

## VI. DISCUSSIONS

In order to look for the viability of extra $U(1)$ superstring inspired model and those of other $E_6$ group emanated ones for doing physics beyond SM, we have investigated the unitarity aspects of these models in detail. For this purpose we have primarily made use of the method due to Durand and Lopez [14] which they have used for flipped $SU(5) \times U(1)$ superstring inspired GUT model. This investigation provides constraints on various SM parameters at weak scale. The variations in the values of different parameters for the $E_6$-$\chi(27)$, $\psi$, $\eta$ and $\nu$ models are negligible except for the case of $\chi(16)$, which occurs due to the significant difference in the one-loop beta functions with respect to those with ***27*** representations. One of the limitations of our work is that for the sake of simplicity, we have not considered the effect of gauge kinetic mixing in these calculations. But this aspect can be explored with more generalization which could further lead not only to "new physics' but in addition may decide the survival of leptophobic models[d] and their usefulness in low energy physics. An interesting outcome of our study is that the masses of the additional neutral gauge boson $Z'$ and *D*-quark are dependent on each other. *D*-quark may decay into a leptoquark or a di-quark [44]. Theoretical calculations about the production of leptoquark with a exchange of $Z'$ in a $e^+e^-$ collision at TESLA energies are under progress which may reveal a rich source of new physics for the ongoing future colliders.

---

[d] To enrich the strong base of leptophobic models, a number of authors [23, 43] have postulated that a possible excess of $Z \to b\bar{b}$ events at CERN $e^+e^-$ collider LEP would be accounted for by the mixing between the $Z$ and leptophobic (hadrophilic) $Z'$. However, from LEP & ALEPH, this probability is this excess is found more weakened.

In Table XIV, we have compared our results of extra $U(1)$ models with those of flipped $SU(5) \times U(1)$ model of Durand and Lopez [14]. We sih to point out that although the flipped $SU(5) \times U(1)$ superstring inspired model have several merits as pointed out by the Durand and Lopez [39] yet the model doesn't have extra $Z'$ and D- quark which occurs naturally in the extra $U(1)$ models considered by us. As such the physics aspect considered by us with the inclusion of $Z'$, was not addressed by Durand and Lopez [14]. Further, It is worth mentioning that a thorough and rigorous investigation about the S,T, U parameters inclusive of triple gauge bosons vertices contributions has been done in the extra $U(1)$ superstring inspired model i.e. $\eta$ model by us [12], which leads to the negative values of the parameters S, T and U. The results obtained by us are in good agreement with experimental signatures. This provides a sound basis for inclusion of the extra $U(1)$ - $\eta$ model as one of the candidates for doing physics beyond SM. We are, however, not aware of any similar calculation done in the context of flipped $SU(5) \times U(1)$ model.

Finally, one may like to speculate as to how the unitarity issue in this model would be different if the supersymmetry is absent? It has been pointed out by the Durand and Lopez [14] and others [32] that the fermionic interaction in theories which are supersymmetric at high energies lead to the same high energy unitarity constraints as scalar interactions. This, therefore, makes it unnecessary to include considerations of fermionic interactions calculations in such theories. But such a situation would not exist in non-supersymmetric theories, implying thereby the inclusion of fermionic interactions calculations in the latter case at high energies. Such an inclusion will obviously complicate the calculations as also the derivations of unitarity constraints.In order to ascertain whether one gets a finite unitarity bound in the case, we are now doing detailed calculations of the same and will report the outcome as soon as the calculation is complete.

## ACKNOWLEDGEMENTS

This work has been fully financially supported by the Department of Science & Technology (DST), New Delhi, India. The authors wish to express their gratitude to DST for this financial support.

## APPENDIX

In Section II, we have used various modified couplings [9, 10] in calculating the amplitudes for the processes $W_L^+ W_L^- \to Z_{1L} Z_{1L}$, $W_L^+ W_L^- \to Z_{2L} Z_{2L}$, $W_L^+ W_L^- \to Z_{1L} Z_{2L}$, and $W_{iL}^{\pm} Z_{iL} \to W_{iL}^{\pm} Z_{iL}$ ($i$ =1,2). The modified couplings are summarized below:

(i) $Z_1 Z_1 H_3^0$:

$$\left[ \frac{1}{2} g_z M_Z \cos\beta \left\{ \cos^2\theta_E + \sin^2\theta_E x_W \tan^2\chi + \sin 2\theta_E \sqrt{x_W} \tan\chi + \frac{\sqrt{x_W}}{3} C_1 \sin 2\theta_E \sec\chi + \frac{2}{3}\sqrt{x_W} C_1 \sin^2\theta_E \sqrt{x_W} \tan\chi \sec\chi + \frac{1}{9} x_W C_1^2 \sin^2\theta_E \sec^2\chi \right\} U_{13} + \frac{1}{2} g_z M_Z \sin\beta \left\{ \cos^2\theta_E + \sin^2\theta_E x_W \tan^2\chi + \sin 2\theta_E \sqrt{x_W} \tan\chi - \frac{4}{3}\sqrt{x_W}\, C_1 \sin 2\theta_E - \frac{8}{3} C_1 \sin^2\theta_E x_W \tan\chi \sec\chi + \frac{16}{9} x_W C_1^2 \sin^2\theta_E \sec^2\chi \right\} U_{23} + \frac{25}{18}\frac{x}{\upsilon} x_W g_z M_Z \sin^2\theta_E \sec^2\chi\, U_{33} \right],$$

(ii) $Z_2 Z_2 H_3^0$:

$$C_1^2 \left[ \frac{1}{2} g_z M_Z \cos\beta \left\{ \sin^2\theta_E + \cos^2\theta_E x_W \tan^2\chi - \sin 2\theta_E \sqrt{x_W} \tan\chi - \frac{\sqrt{x_W}}{3} \sin 2\theta_E \sec\chi + \frac{2}{3} \cos^2\theta_E x_W \tan\chi \sec\chi + \frac{1}{9} x_W \sin^2\theta_E \sec^2\chi \right\} U_{13} + \frac{1}{2} g_z M_Z \sin\beta \left\{ \sin^2\theta_E + \cos^2\theta_E x_W \tan^2\chi - \sin 2\theta_E \sqrt{x_W} \tan\chi + \frac{8}{3}\sqrt{x_W} \sin 2\theta_E \sec\chi - \frac{8}{3} \cos^2\theta_E x_W \tan\chi \sec\chi + \frac{16}{9} x_W \cos^2\theta_E \sec^2\chi \right\} U_{23} + \frac{25}{18}\frac{x}{\upsilon} x_W g_z M_Z \cos^2\theta_E \sec^2\chi\, U_{33} \right],$$

(iii) $W_L^+ W_L^- H_3^0$: $\quad g_L M_W \left( \cos\beta\ U_{13} + \sin\beta\ U_{23} \right)$,

(iv) $W^{\pm} H^{\mp} Z_1$: $\quad \frac{5}{6}\left(4\sqrt{2} G_F\right)^{\frac{1}{2}} M_W M_Z \sin\theta_W \sin 2\beta\ \ C_1 \frac{\sin\theta_E}{\cos\chi}$,

(v) (v) $W^{\pm} H^{\mp} Z_2$: $\quad \frac{5}{6}\left(4\sqrt{2} G_F\right)^{\frac{1}{2}} M_W M_Z \sin\theta_W \sin 2\beta\ \ C_1 \frac{\cos\theta_E}{\cos\chi}$.

## TABLE CAPTIONS

**Table I:** The calculated unitarity constrained on $x$ for the process $W_L^+ W_L^- \to Z_{2L} Z_{2L}$ in $E_6$ models, using Low energy theorems.

**Table II:** The matrix T1 with $T_i = \frac{\lambda_i^2}{4\pi}$, where $i = H,t,b,D$. The matrix is symmetric Hermitian.

**Table III:** The matrix T2 with notations identical as in Table II.

**Table IV:** The matrix T3 with notations identical as in Table II.

**Table V:** The matrix T4 with notations identical as in Table II.

**Table VI:** The normalized $U(1)'$ charge, $Q_E'$, of all the matter fields in the **<u>27</u>** representation for the $\chi$, $\psi$, $\eta$, $\nu$ $E_6$ models with the normalization $Q_E' = \sqrt{3/5} Q_E$, where $Q_E$ is unrenormalized charge.

**Table VII:** The unitary bounds obtained on the superpotential couplings at weak scale in the $E_6$ extra $U(1)$ models, when the contribution of $\lambda_b$ at GUT scale is taken to the zero i.e. $\lambda_t \neq \lambda_b$ at GUT scale.

**Table VIII:** The bounds on $\lambda_t, \lambda_b$ at weak scale in $E_6$ extra $U(1)$ models, with the assumption $\lambda_t = \lambda_b$ at GUT scale.

**Table IX:** Critical bounds on the couplings at the weak scale using analytical solution.

**Table X:** The estimated upper bounds on top quark in $E_6$ extra $U(1)$ models using unitarity constraint on $\lambda_t$ for the cases $\lambda_t \neq \lambda_b$ and $\lambda_t = \lambda_b$ at GUT scale.

**Table XI:** The upper bounds on $\tan\beta$ calculated in $E_6$ extra $U(1)$ models for the presently known experimental values of on top quark mass.

**Table XII:** Lower bounds on $\tan\beta$ for the $E_6$ extra $U(1)$ models for $m_b = 4.5\ GeV$.

**Table XIII:** Theoretical bounds on lightest neutral scalar Higgs mass in $E_6$ extra $U(1)$ models pertaining to the measured value of top quark mass at CDF and D∅. The bounds listed in first row correspond for the case $\lambda_t \neq \lambda_b$ and in second row for $\lambda_t = \lambda_b$ at GUT respectively.

**Table XIV:** A comparison between the extra $U(1)$ models with those of flipped $SU(5)\times U(1)$ model for the bounds on various parameters obtained through our analysis for the case $\lambda_t \neq \lambda_b$ at GUT.

## FIGURE CAPTIONS

**Fig. 1(a,b,c):** Tree level Feynman diagrams contributing to $M_{gauge}$ part of the amplitude for the process $W_L^+ W_L^- \to Z_L Z_L$.

**Fig. 2:** s- channel tree level Feynman diagram contributing to $M_{SB}$ part of the amplitude for the process $W_L^+ W_L^- \to Z_L Z_L$.

**Fig. 3(a,b):** t- channel and u- channel Feynman diagrams contributing to $M_{SB}$ part of the amplitude for the process $W_L^+ W_L^- \to Z_{iL} Z_{jL}$ $(i, j = 1,2)$ in the extra $U(1)$ superstring inspired model.

**Fig. 4:** The unitarity surface in the subspace of the scalar superpotential couplings $\lambda_t, \lambda_H, \lambda_D$ for $\lambda_b = 0$ for the energy range $M_Z << \sqrt{s} \leq M_G$.

**Fig. 5:** The unitarity surface of Yukawa couplings at weak scale ($M_Z$) obtained by evolving the surface at ($M_G$) shown in Fig. 4 from GUT scale to weak scale for the case of $E_6$-$\eta$ model $(\beta_E \cong 9.64)$.

**Fig. 6:** (a) The projection of unitarity boundaries are shown in the plane $\lambda_t, \lambda_H$ plane at GUT scale. This leads to unitarity constrained region of Yukawa couplings at weak scale are shown in (b) $\chi(16)$ $(\beta_E \cong 6.023)$, (c) $\chi(27)$, $(\beta_E \cong 9.023)$, (d) $\psi$ $(\beta_E \cong 9.028)$, (e) $\eta$ $(\beta_E \cong 9.64)$ and (f) $\nu$ $(\beta_E \cong 9.9)$ $E_6$ models. The effect of evolution for $\lambda_t \neq \lambda_b (\lambda_b = 0)$ and $\lambda_t = \lambda_b$ are represented by curve "(i)" and curve "(ii)" respectively.

**Fig. 7(a):** The possible values of $x/v$ for the favored values of $\tan\beta$ of $E_6$ extra $U(1)$ models for various $Z_2$ masses.

**Fig. 7(b):** A plausible range for *D*-quark mass against for various $M_{Z_2}$ .

**Table I**

| $\theta_{E_6}$ (Model) | $0\,(\chi)$ | $\pi/2\,(\psi)$ | $\tan^{-1}(-\sqrt{5/3})\,(\eta)$ | $\tan^{-1}(-\sqrt{15})\,(\nu)$ |
|---|---|---|---|---|
| $\boldsymbol{x}$ | $\leq 2040\, GeV$ | $\leq 2040\, GeV$ | $\leq 2883\, GeV$ | $\leq 2607\, GeV$ |

**Table II**

$$
\begin{bmatrix}
\frac{T_H}{4} \\
\frac{T_H}{8\sqrt{2}} & \frac{3T_H}{16} \\
-\frac{T_H}{4\sqrt{2}} & -\frac{9T_H}{8} & 0 \\
0 & \frac{T_H}{8\sqrt{2}} & -\frac{T_H}{4\sqrt{2}} & \frac{T_H}{4} \\
\frac{T_H}{8\sqrt{2}} & \frac{T_H}{16} & -\frac{3T_H}{16} & \frac{T_H}{8\sqrt{2}} & \frac{T_H}{16} \\
-\frac{T_H}{8\sqrt{2}} & \frac{T_H}{16} & -\frac{3T_H}{16} & -\frac{T_H}{8\sqrt{2}} & -\frac{T_H}{16} & \frac{T_H}{32} \\
\frac{T_H}{8\sqrt{2}} & 0 & -\frac{3T_H}{16} & \frac{T_H}{8\sqrt{2}} & \frac{T_H}{16} & \frac{T_H}{16} & \frac{3T_H}{32} \\
-\frac{T_H}{8\sqrt{2}} & \frac{3T_H}{16} & 0 & -\frac{T_H}{16\sqrt{2}} & \frac{3T_H}{16} & \frac{3T_H}{32} & \frac{3T_H}{32} & 0 \\
0 & -\frac{T_H}{8\sqrt{2}} & 0 & 0 & -\frac{T_H}{16\sqrt{2}} & \frac{T_H}{16\sqrt{2}} & -\frac{T_H}{16\sqrt{2}} & 0 & -\frac{T_H}{16} \\
0 & -\frac{T_H}{8\sqrt{2}} & 0 & 0 & -\frac{T_H}{16\sqrt{2}} & \frac{T_H}{16\sqrt{2}} & -\frac{T_H}{16\sqrt{2}} & 0 & 0 & -\frac{T_H}{16} \\
0 & 0 & -\frac{T_H}{8\sqrt{2}} & 0 & 0 & 0 & 0 & \frac{T_H}{32\sqrt{2}} & 0 & 0 & \frac{T_H}{4} \\
0 & 0 & 0 & 0 & 0 & 0 & 0 & \frac{T_H}{32\sqrt{2}} & 0 & 0 & -\frac{T_H}{8} & -\frac{T_H}{8} \\
\frac{T_b}{8} & \frac{T_b}{8\sqrt{2}} & 0 & \frac{T_b}{8} & \frac{T_b}{8\sqrt{2}} & \frac{T_b}{8\sqrt{2}} & \frac{T_b}{8\sqrt{2}} & 0 & -\frac{T_b}{8} & -\frac{T_b}{8} & -\frac{T_b}{8} & -\frac{T_b}{8} & 0 \\
0 & \frac{T_b}{8\sqrt{2}} & 0 & 0 & \frac{T_b}{8\sqrt{2}} & \frac{T_b}{8\sqrt{2}} & \frac{T_b}{8\sqrt{2}} & 0 & 0 & 0 & -\frac{T_b}{8} & -\frac{T_b}{8} & 0 & 0 \\
\frac{T_t}{8} & \frac{T_t}{8\sqrt{2}} & 0 & \frac{T_t}{8} & \frac{T_t}{8\sqrt{2}} & \frac{T_t}{8\sqrt{2}} & \frac{T_t}{8\sqrt{2}} & 0 & \frac{T_t}{8} & \frac{T_t}{8} & \frac{T_t}{8} & \frac{T_t}{8} & 0 & 0 & 0 \\
0 & \frac{T_t}{8\sqrt{2}} & 0 & 0 & \frac{T_t}{8\sqrt{2}} & \frac{T_t}{8\sqrt{2}} & \frac{T_t}{8\sqrt{2}} & 0 & 0 & 0 & \frac{T_t}{8} & \frac{T_t}{8} & 0 & 0 & 0 & 0 \\
\frac{\sqrt{T_D T_H}}{4} & \frac{\sqrt{T_D T_H}}{4\sqrt{2}} & 0 & -\frac{\sqrt{T_D T_H}}{4} & -\frac{\sqrt{T_D T_H}}{4\sqrt{2}} & \frac{\sqrt{T_D T_H}}{4\sqrt{2}} & -\frac{\sqrt{T_D T_H}}{4\sqrt{2}} & 0 & \frac{\sqrt{T_D T_H}}{4} & -\frac{\sqrt{T_D T_H}}{4} & 0 & 0 & 0 & 0 & 0 & 0 & T_D \\
\frac{T_H}{8} & 0 & 0 & \frac{T_H}{8} & 0 & 0 & 0 & 0 & -\frac{T_H}{8} & -\frac{T_H}{8} & -\frac{\iota T_H}{8} & 0 & 0 & 0 & 0 & 0 & -\frac{\iota\sqrt{T_D T_H}}{4} & 0 \\
-\frac{\iota T_H}{8} & 0 & 0 & -\frac{\iota T_H}{8} & 0 & 0 & 0 & 0 & -\frac{\iota T_H}{8} & -\frac{T_H}{8} & 0 & 0 & 0 & 0 & 0 & 0 & \frac{\iota\sqrt{T_D T_H}}{4} & 0 & 0
\end{bmatrix}
$$

**Table III**

$$\begin{bmatrix}
T_bT_H & T_bT_H + T_tT_H & \frac{\sqrt{T_bT_H}}{\sqrt{2}} & \frac{\imath\sqrt{T_bT_H}}{\sqrt{2}} & \frac{\sqrt{T_bT_H}}{\sqrt{2}} & \frac{\imath\sqrt{T_bT_H}}{\sqrt{2}} & \frac{\imath\sqrt{T_bT_H}}{\sqrt{2}} & -\frac{\sqrt{T_bT_H}}{\sqrt{2}} & \frac{\imath\sqrt{T_bT_H}}{\sqrt{2}} & \imath\sqrt{2T_bT_H} \\
T_bT_H + T_tT_H & T_tT_H & -\frac{\sqrt{T_tT_H}}{\sqrt{2}} & \frac{\imath\sqrt{T_tT_H}}{\sqrt{2}} & \frac{\sqrt{T_tT_H}}{\sqrt{2}} & \frac{\imath\sqrt{T_tT_H}}{\sqrt{2}} & \frac{\imath\sqrt{T_tT_H}}{\sqrt{2}} & -\frac{\sqrt{T_tT_H}}{\sqrt{2}} & \frac{\imath\sqrt{T_tT_H}}{\sqrt{2}} & \imath\sqrt{2T_tT_H} \\
\frac{\sqrt{T_bT_H}}{\sqrt{2}} & -\frac{\sqrt{T_tT_H}}{\sqrt{2}} & 0 & 0 & 0 & 0 & 0 & 0 & 0 & 0 \\
-\frac{\imath\sqrt{T_bT_H}}{\sqrt{2}} & -\frac{\imath\sqrt{T_tT_H}}{\sqrt{2}} & 0 & 0 & 0 & 0 & 0 & 0 & 0 & 0 \\
\frac{\sqrt{T_bT_H}}{\sqrt{2}} & \frac{\sqrt{T_tT_H}}{\sqrt{2}} & 0 & 0 & 0 & 0 & 0 & 0 & 0 & 0 \\
-\frac{\imath\sqrt{T_bT_H}}{\sqrt{2}} & -\frac{\imath\sqrt{T_tT_H}}{\sqrt{2}} & 0 & 0 & 0 & 0 & 0 & 0 & 0 & 0 \\
-\frac{\imath\sqrt{T_bT_H}}{\sqrt{2}} & -\frac{\imath\sqrt{T_tT_H}}{\sqrt{2}} & 0 & 0 & 0 & 0 & 0 & 0 & 0 & 0 \\
-\frac{\sqrt{T_bT_H}}{\sqrt{2}} & -\frac{\sqrt{T_tT_H}}{\sqrt{2}} & 0 & 0 & 0 & 0 & 0 & 0 & 0 & 0 \\
-\frac{\imath\sqrt{T_bT_H}}{\sqrt{2}} & -\frac{\imath\sqrt{T_tT_H}}{\sqrt{2}} & 0 & 0 & 0 & 0 & 0 & 0 & 0 & 0 \\
-\imath\sqrt{2T_bT_H} & -\imath\sqrt{2T_tT_H} & 0 & 0 & 0 & 0 & 0 & 0 & 0 & 0
\end{bmatrix}$$

**Table IV**

$$\begin{bmatrix}
0 & \frac{\sqrt{T_HT_t}}{\sqrt{2}} & -\frac{\sqrt{T_HT_t}}{\sqrt{2}} & \frac{\imath\sqrt{T_HT_t}}{\sqrt{2}} & -\imath\sqrt{2T_DT_H} \\
\frac{\sqrt{T_HT_t}}{\sqrt{2}} & 0 & 0 & 0 & 0 \\
-\frac{\sqrt{T_HT_t}}{\sqrt{2}} & 0 & 0 & 0 & 0 \\
-\frac{\imath\sqrt{T_HT_t}}{\sqrt{2}} & 0 & 0 & 0 & 0 \\
\imath\sqrt{2T_DT_H} & 0 & 0 & 0 & 0
\end{bmatrix}$$

**Table V**

$$\begin{bmatrix}
0 & \frac{\sqrt{T_bT_H}}{\sqrt{2}} & \frac{\sqrt{T_bT_H}}{\sqrt{2}} & \frac{\imath\sqrt{T_bT_H}}{\sqrt{2}} & \imath\sqrt{2T_DT_H} \\
\frac{\sqrt{T_bT_H}}{\sqrt{2}} & 0 & 0 & 0 & 0 \\
\frac{\sqrt{T_bT_H}}{\sqrt{2}} & 0 & 0 & 0 & 0 \\
-\frac{\imath\sqrt{T_bT_H}}{\sqrt{2}} & 0 & 0 & 0 & 0 \\
-\imath\sqrt{2T_DT_H} & 0 & 0 & 0 & 0
\end{bmatrix}$$

**Table VI**

| ***SO(10)*** | ***SU(5)*** | **Fields** | $2\sqrt{6}\,Q_\chi$ | $Q'_\chi$ | $\sqrt{72/5}\,Q_\psi$ | $Q'_\psi$ | $Q_\eta$ | $Q'_\eta$ | $Q_\nu$ | $Q'_\nu$ |
|---|---|---|---|---|---|---|---|---|---|---|
| **16** | **10** | $Q$ | -1 | $-1/2\sqrt{10}$ | +1 | $1/2\sqrt{6}$ | -1/3 | $-\sqrt{1/15}$ | $\sqrt{1/24}$ | $1/2\sqrt{10}$ |
| | | $u^c$ | -1 | $-1/2\sqrt{10}$ | +1 | $1/2\sqrt{6}$ | -1/3 | $-\sqrt{1/15}$ | $\sqrt{1/24}$ | $1/2\sqrt{10}$ |
| | | $e^c$ | -1 | $-1/2\sqrt{10}$ | +1 | $1/2\sqrt{6}$ | -1/3 | $-\sqrt{1/15}$ | $\sqrt{1/24}$ | $1/2\sqrt{10}$ |
| | $\bar{5}$ | $L$ | +3 | $3/2\sqrt{10}$ | +1 | $1/2\sqrt{6}$ | +1/6 | $1/2\sqrt{15}$ | $\sqrt{1/6}$ | $1/\sqrt{10}$ |
| | | $d^c$ | +3 | $3/2\sqrt{10}$ | +1 | $1/2\sqrt{6}$ | +1/6 | $1/2\sqrt{15}$ | $\sqrt{1/6}$ | $1/\sqrt{10}$ |
| | **1** | $\nu^c$ | -5 | $5/2\sqrt{10}$ | +1 | $1/2\sqrt{6}$ | -5/6 | $-\sqrt{5}/2\sqrt{3}$ | 0 | 0 |
| **10** | **5** | $H_u$ | +2 | $1/\sqrt{10}$ | -2 | $-1/\sqrt{6}$ | +2/3 | $2/\sqrt{15}$ | $-\sqrt{1/6}$ | $-1/\sqrt{10}$ |
| | | $D$ | +2 | $1/\sqrt{10}$ | -2 | $-1/\sqrt{6}$ | +2/3 | $2/\sqrt{15}$ | $-\sqrt{1/6}$ | $-1/\sqrt{10}$ |
| | $\bar{5}$ | $H_d$ | -2 | $-1/\sqrt{10}$ | -2 | $-1/\sqrt{6}$ | +1/6 | $1/2\sqrt{15}$ | $-\sqrt{3/8}$ | $-3/2\sqrt{10}$ |
| | | $\overline{D}$ | -2 | $-1/\sqrt{10}$ | -2 | $-1/\sqrt{6}$ | +1/6 | $1/2\sqrt{15}$ | $-\sqrt{3/8}$ | $-3/2\sqrt{10}$ |
| **1** | **1** | $S$ | 0 | 0 | 4 | $2/\sqrt{6}$ | -5/6 | $-\sqrt{5}/2\sqrt{3}$ | $\sqrt{25/24}$ | $5\sqrt{5}/8\sqrt{3}$ |

**Table VII**

| $E_6$ | $b_1$ | $b_2$ | $b_3$ | $b_E$ | $g_1$ | $g_2$ | $g_3$ | $g_E$ | $\lambda_t$ | $\lambda_b$ | $\lambda_H$ | $\lambda_D$ |
|---|---|---|---|---|---|---|---|---|---|---|---|---|
| $\chi$ (16) | 6.6 | 1 | -3 | 6.023<br>6.01 | 0.540 | 0.957 | 1.350 | 0.5596<br>0.5601 | 1.466 | 1.464 | 1.057 | 1.392 |
| $\chi$ (27) | 9.6 | 4 | 0 | 9.023<br>9.01 | 0.462 | 0.653 | 1.218 | 0.4742<br>0.4744 | 1.292 | 1.282 | 0.929 | 1.293 |
| $\psi$ | 9.6 | 4 | 0 | 9.007<br>9.028<br>- 9.028 | 0.462 | 0.653 | 1.218 | 0.4745<br>0.4740<br>0.4740 | 1.290 | 1.284 | 0.975 | 1.326 |
| $\eta$ | 9.6 | 4 | 0 | 9.04<br>9.64 | 0.462 | 0.652 | 1.218 | 0.4738<br>0.4610 | 1.295<br>1.294 | 1.274<br>1.278 | 0.957<br>0.955 | 1.314<br>1.312 |
| $\nu$ | 9.6 | 4 | 0 | 9.24<br>9.9 | 0.462 | 0.653 | 1.218 | 0.4694<br>0.4557 | 1.287<br>1.286 | 1.287<br>1.286 | 0.971<br>0.968 | 1.323<br>1.321 |

**Table VIII**

| $E_6$ | $\lambda'_t$ | $\lambda'_b$ |
|---|---|---|
| $\chi$ (16) | 1.353 | 1.355 |
| $\chi$ (27) | 1.197 | 1.196 |
| $\psi$ | 1.196 | 1.187 |
| $\eta$ | 1.202 | 1.180 |
| $\nu$ | 1.191 | 1.191 |

**Table IX**

| $E_6$ | $b_1$ | $b_2$ | $b_3$ | $b_E$ | $g_1$ | $g_2$ | $g_3$ | $g_E$ | $\lambda_{t,c}$ | $\lambda_{b,c}$ | $\lambda_{H,c}$ | $\lambda_{D,c}$ |
|---|---|---|---|---|---|---|---|---|---|---|---|---|
| $\chi$ (16) | 6.6 | 1 | -3 | 6.023<br>6.01 | 0.540 | 0.957 | 1.350 | 0.5596<br>0.5601 | 1.481 | 1.480 | 1.071 | 1.423 |
| $\chi$ (27) | 9.6 | 4 | 0 | 9.023<br>9.01 | 0.462 | 0.653 | 1.218 | 0.4742<br>0.4744 | 1.292 | 1.288 | 0.944 | 1.295 |
| $\psi$ | 9.6 | 4 | 0 | 9.007<br>9.028<br>- 9.028 | 0.462 | 0.653 | 1.218 | 0.4745<br>0.4740<br>0.4740 | 1.291 | 1.284 | 0.961 | 1.328 |
| $\eta$ | 9.6 | 4 | 0 | 9.04<br>9.64 | 0.462 | 0.652 | 1.218 | 0.4738<br>0.4610 | 1.295<br>1.295 | 1.279<br>1.279 | 0.970<br>0.968 | 1.316<br>1.315 |
| $\nu$ | 9.6 | 4 | 0 | 9.24<br>9.9 | 0.462 | 0.653 | 1.218 | 0.4694<br>0.4557 | 1.287<br>1.287 | 1.287<br>1.287 | 0.983<br>0.980 | 1.325<br>1.323 |

**Table X**

| Estimated | when $\lambda_t \neq \lambda_b$ at GUT | | | | | when $\lambda_t = \lambda_b$ at GUT | | | | |
|---|---|---|---|---|---|---|---|---|---|---|
| | $\chi$ (16) | $\chi$ (27) | $\psi$ | $\eta$ | $\nu$ | $\chi$ (16) | $\chi$ (27) | $\psi$ | $\eta$ | $\nu$ |
| $m_t$ (GeV) | 255 | 225 | 224 | 225 | 224 | 235 | 208 | 208 | 209 | 207 |

**Table XI**

| $m_t$ **(GeV)** | $\tan\beta$ **, when** $\lambda_t \neq \lambda_b$ **at GUT** | | | | | $\tan\beta$ **, when** $\lambda_t = \lambda_b$ **at GUT** | | | | |
|---|---|---|---|---|---|---|---|---|---|---|
| | $\chi$ (16) | $\chi$ (27) | $\psi$ | $\eta$ | $\nu$ | $\chi$ (16) | $\chi$ (27) | $\psi$ | $\eta$ | $\nu$ |
| **167.4±11.4 [CDF (*Di-l*)]** | 1.150 | 0.896 | 0.893 | 0.901 | 0.889 | 0.989 | 0.740 | 0.736 | 0.749 | 0.730 |
| **168.4±12.8 [D∅ (*Di- l*)]** | 1.138 | 0.884 | 0.881 | 0.889 | 0.877 | 0.976 | 0.728 | 0.726 | 0.737 | 0.717 |
| **176.1±7.3 [CDF (*l+j*)]** | 1.048 | 0.794 | 0.790 | 0.798 | 0.786 | 0.887 | 0.632 | 0.630 | 0.641 | 0.620 |
| **180.1±5.3 [D∅ (*l+j*)]** | 1.003 | 0.747 | 0.744 | 0.752 | 0.739 | 0.842 | 0.581 | 0.579 | 0.590 | 0.570 |
| **186±11.5 [D∅ (*All jets*)]** | 0.939 | 0.679 | 0.676 | 0.684 | 0.670 | 0.776 | 0.504 | 0.502 | 0.514 | 0.491 |
| **178±4.3 Average [Tevatron Run I]** | 1.026 | 0.771 | 0.768 | 0.776 | 0.763 | 0.886 | 0.608 | 0.606 | 0.617 | 0.596 |

**Table XII**

| $m_b$ (GeV) | $\tan\beta$ , when $\lambda_t \neq \lambda_b$ at GUT | | | | | $\tan\beta$ , when $\lambda_t = \lambda_b$ at GUT | | | | |
|---|---|---|---|---|---|---|---|---|---|---|
| | $\chi$ (16) | $\chi$ (27) | $\psi$ | $\eta$ | $\nu$ | $\chi$ (16) | $\chi$ (27) | $\psi$ | $\eta$ | $\nu$ |
| **4.5** | 0.018 | 0.020 | 0.020 | 0.020 | 0.020 | 0.019 | 0.021 | 0.021 | 0.022 | 0.022 |

**Table XIII**

| $m_t$ (GeV) | $m_{H_2^0}^{tree}$ $(GeV)$ at tree level | | | | | $m_{H_2^0}$ $(GeV)$ after one-loop stop (2 *TeV*) contribution | | | | |
|---|---|---|---|---|---|---|---|---|---|---|
| | $\chi$ (16) | $\chi$ (27) | $\psi$ | $\eta$ | $\nu$ | $\chi$ (16) | $\chi$ (27) | $\psi$ | $\eta$ | $\nu$ |
| **167.4±11.4 [CDF (*Di-l*)]** | 187<br>187 | 165<br>160 | 172<br>167 | 169<br>165 | 171<br>166 | 238<br>238 | 221<br>217 | 226<br>222 | 224<br>221 | 225<br>222 |
| **168.4±12.8 [D∅ (*Di- l*)]** | 187<br>187 | 164<br>159 | 172<br>166 | 169<br>164 | 171<br>166 | 239<br>239 | 221<br>217 | 227<br>222 | 224<br>221 | 226<br>222 |
| **176.1±7.3 [CDF (*l*+*j*)]** | 188<br>186 | 162<br>154 | 169<br>160 | 167<br>158 | 169<br>159 | 243<br>241 | 223<br>217 | 228<br>222 | 227<br>220 | 228<br>221 |
| **180.1±5.3 [D∅ (*l*+*j*)]** | 188<br>185 | 160<br>150 | 167<br>156 | 165<br>154 | 166<br>155 | 245<br>242 | 224<br>217 | 229<br>221 | 227<br>220 | 228<br>220 |
| **186±11.5 [D∅ (*All jets*)]** | 187<br>182 | 157<br>143 | 164<br>148 | 161<br>147 | 163<br>147 | 247<br>243 | 225<br>216 | 230<br>218 | 228<br>218 | 229<br>218 |
| **178±4.3 Average [Tevatron Run I]** | 188<br>186 | 161<br>152 | 168<br>158 | 169<br>159 | 167<br>157 | 244<br>242 | 223<br>217 | 229<br>221 | 229<br>222 | 228<br>220 |

## Table XIV

| Models | $\lambda_t$ | $\lambda_b$ | $\lambda_H$ | $\lambda_D$ | $m_t\ (GeV)$ | $m_{H_2^0}^{tree}\ (GeV)$ | $m_{H_2^0}\ GeV$ |
|---|---|---|---|---|---|---|---|
| $E_6$-$\chi$ (16) | 1.466 | 1.464 | 1.057 | 1.392 | 255 | 188 | 244 |
| $E_6$-$\chi$ (27) | 1.292 | 1.282 | 0.929 | 1.293 | 225 | 161 | 223 |
| $E_6$-$\psi$ | 1.290 | 1.284 | 0.975 | 1.326 | 224 | 168 | 229 |
| $E_6$-$\eta$ | $\left.\begin{matrix}1.295\\1.294\end{matrix}\right\}$ | $\left.\begin{matrix}1.274\\1.278\end{matrix}\right\}$ | $\left.\begin{matrix}0.957\\0.955\end{matrix}\right\}$ | $\left.\begin{matrix}1.314\\1.312\end{matrix}\right\}$ | 225 | 169 | 229 |
| $E_6$-$\nu$ | $\left.\begin{matrix}1.287\\1.286\end{matrix}\right\}$ | $\left.\begin{matrix}1.287\\1.286\end{matrix}\right\}$ | $\left.\begin{matrix}0.971\\0.968\end{matrix}\right\}$ | $\left.\begin{matrix}1.323\\1.321\end{matrix}\right\}$ | 224 | 167 | 228 |
| flipped $SU(5)\times U(1)$ | 1.286 | 1.277 | 0.921 | - | 224 | 158 | 221 |

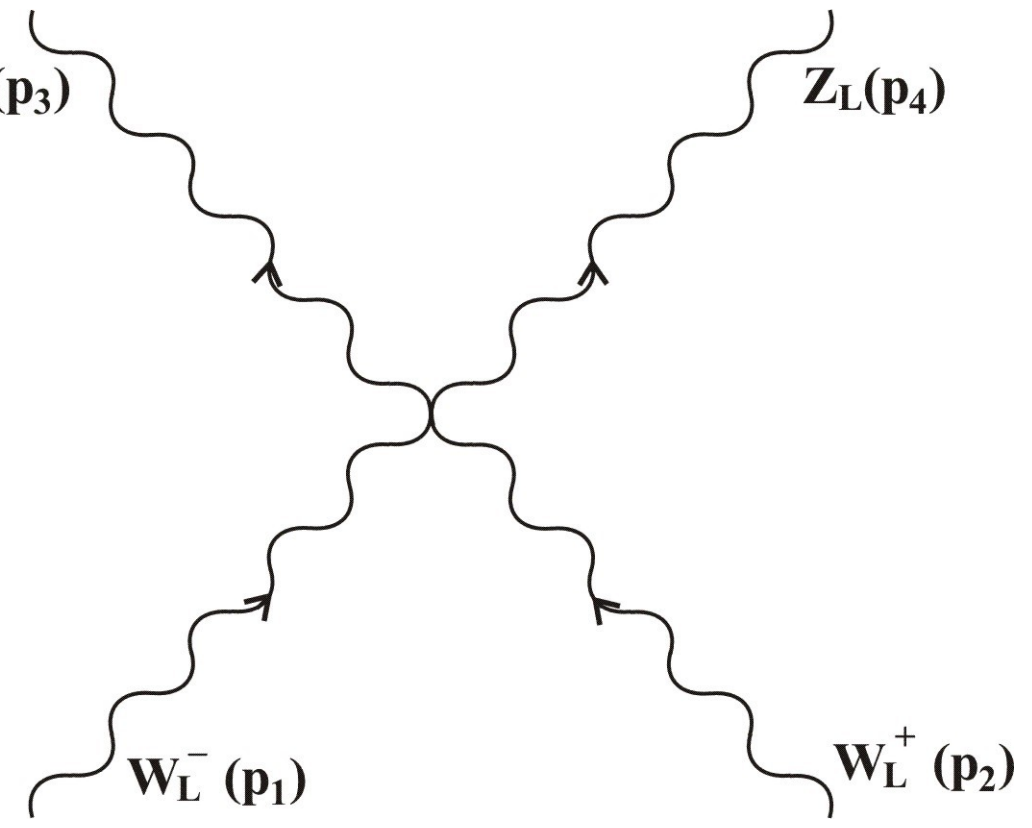

Fig. 1(a)

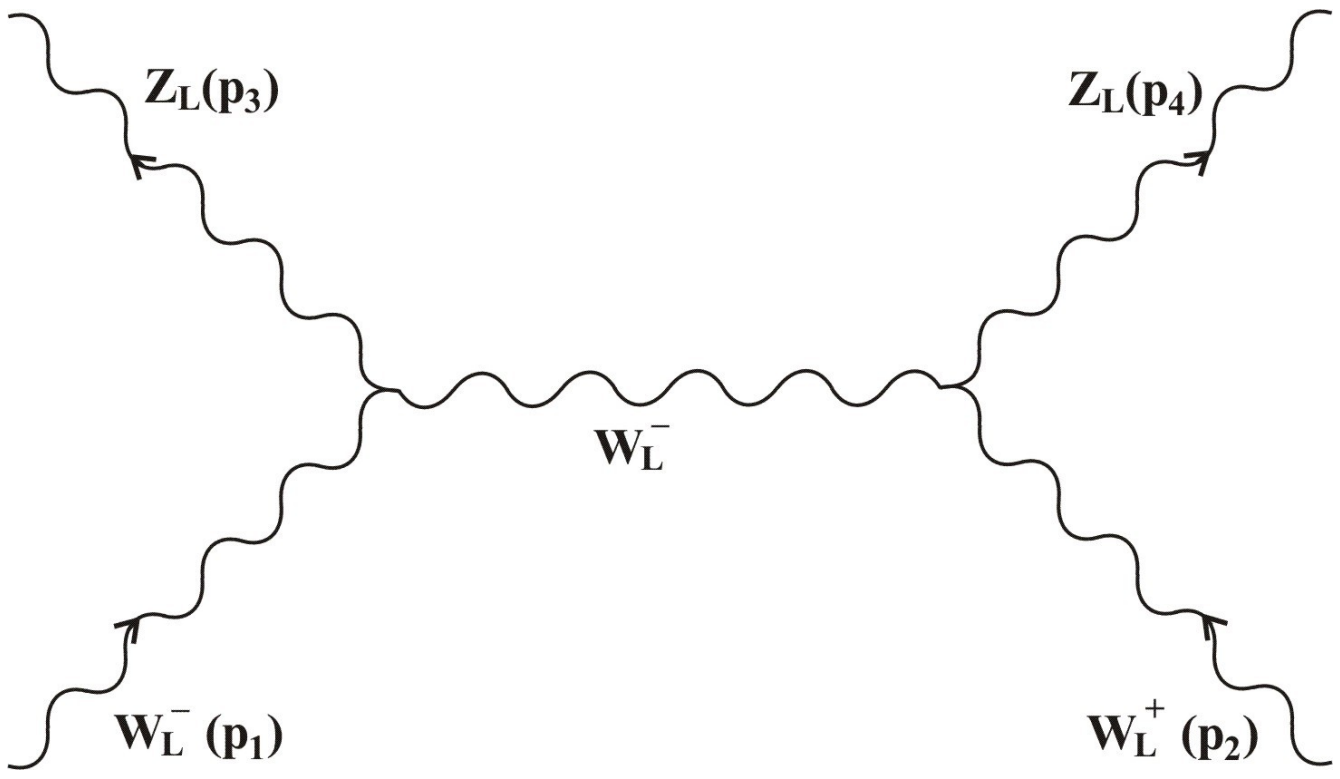

Fig. 1(b)

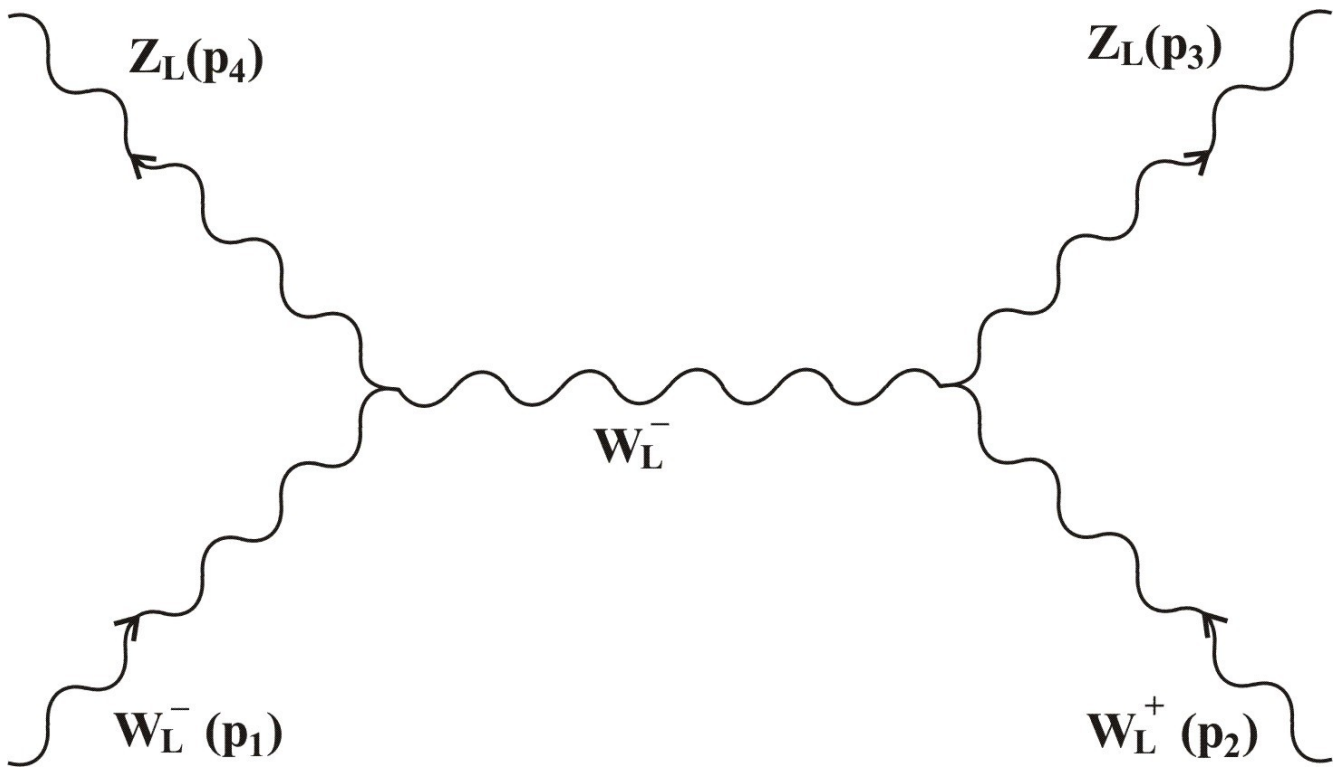

Fig. 1(c)

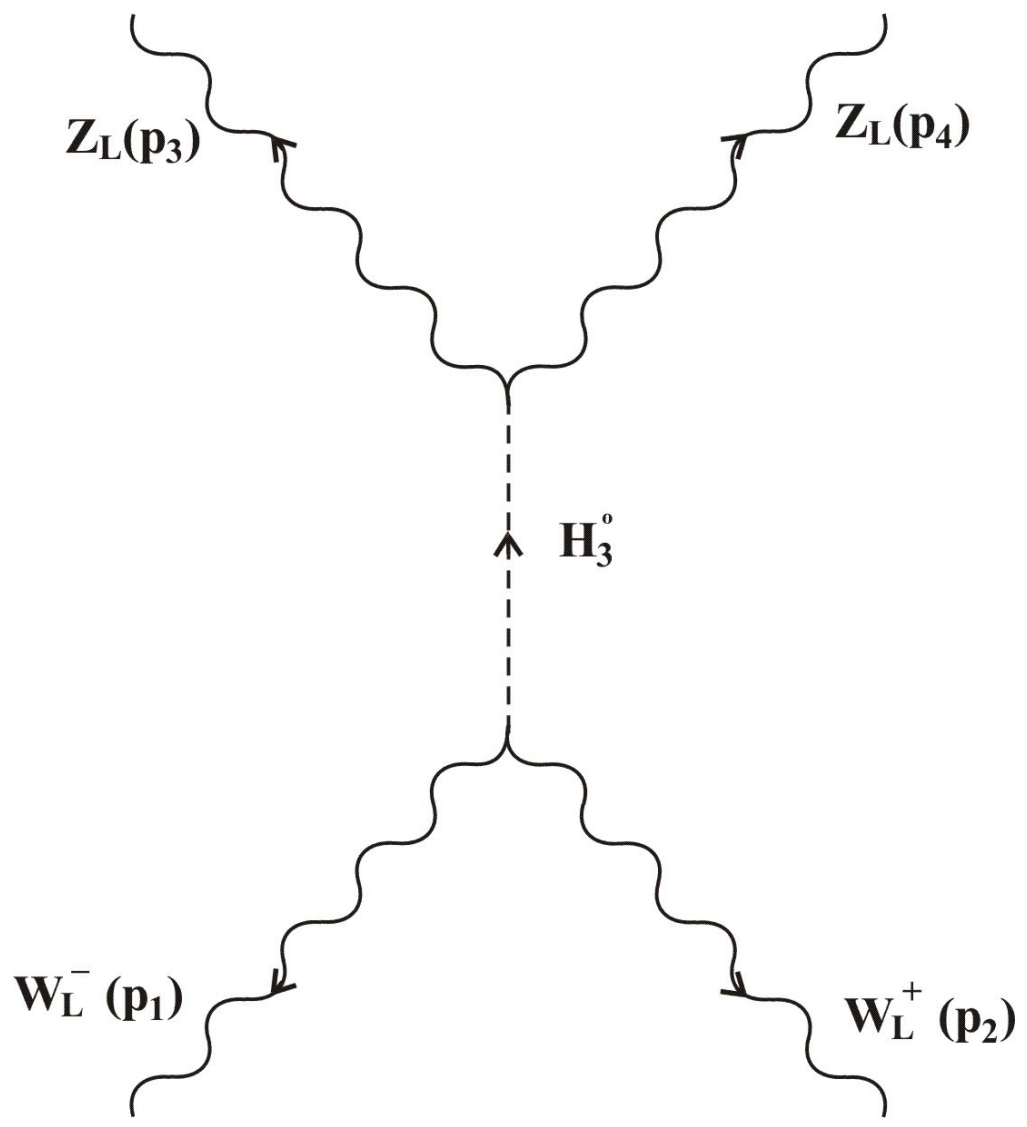

Fig. 2

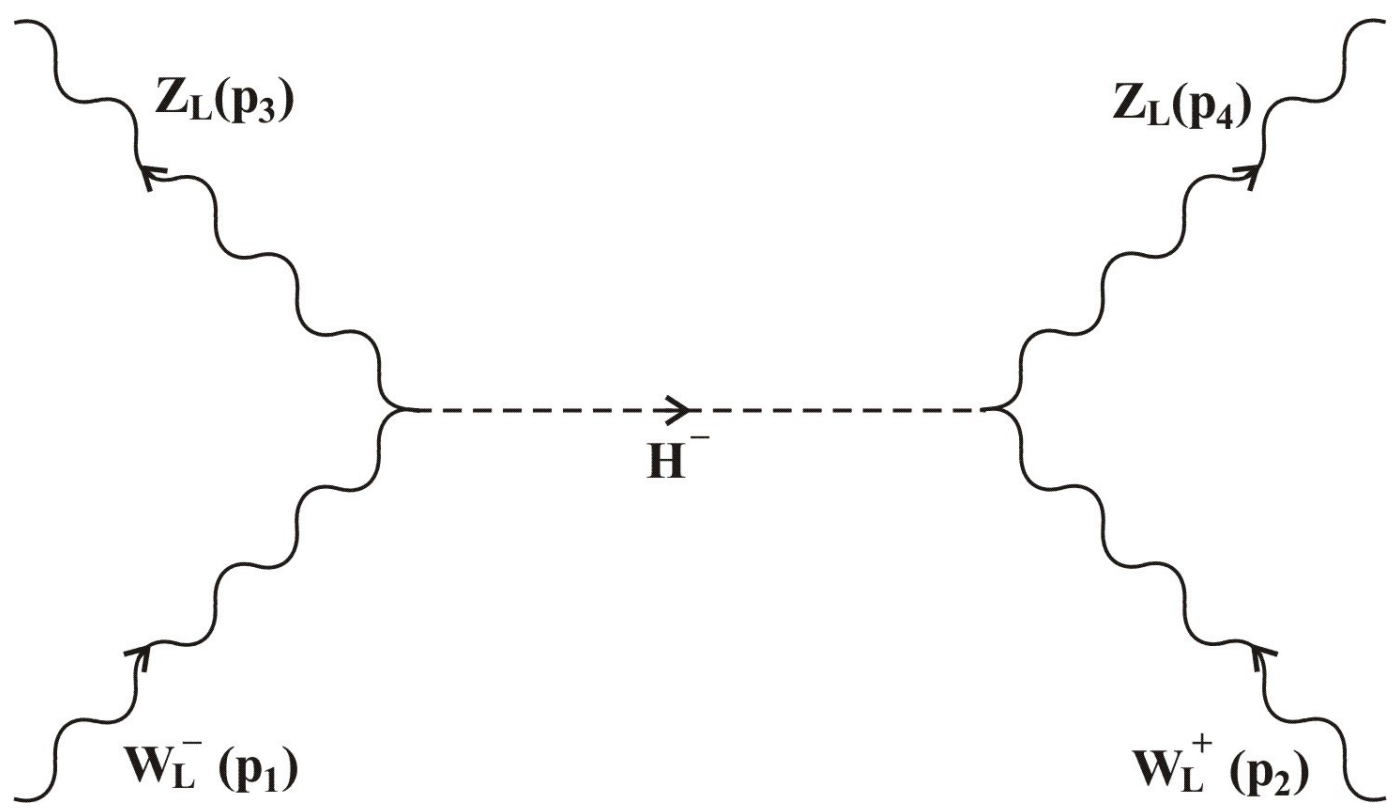

Fig. 3(a)

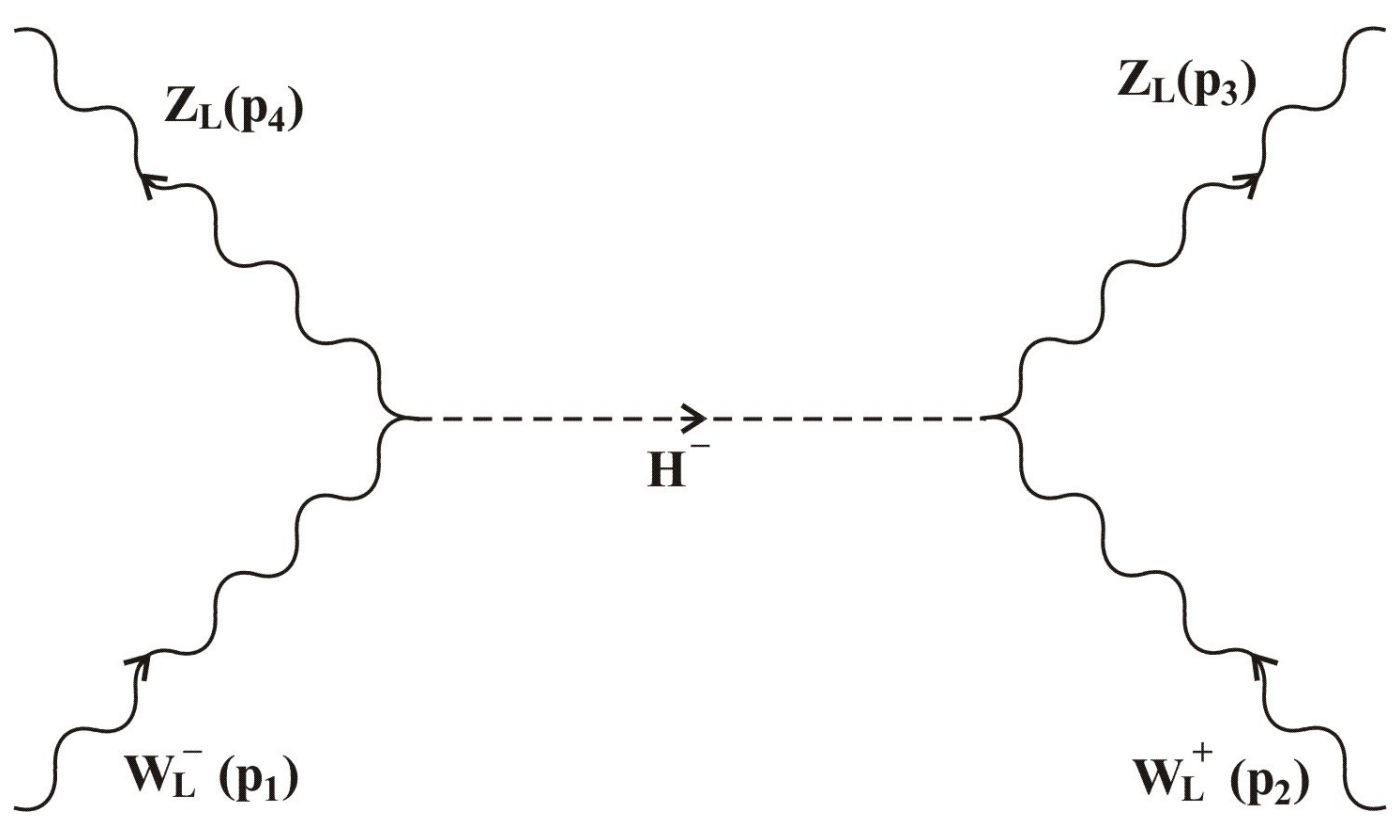

Fig. 3(b)

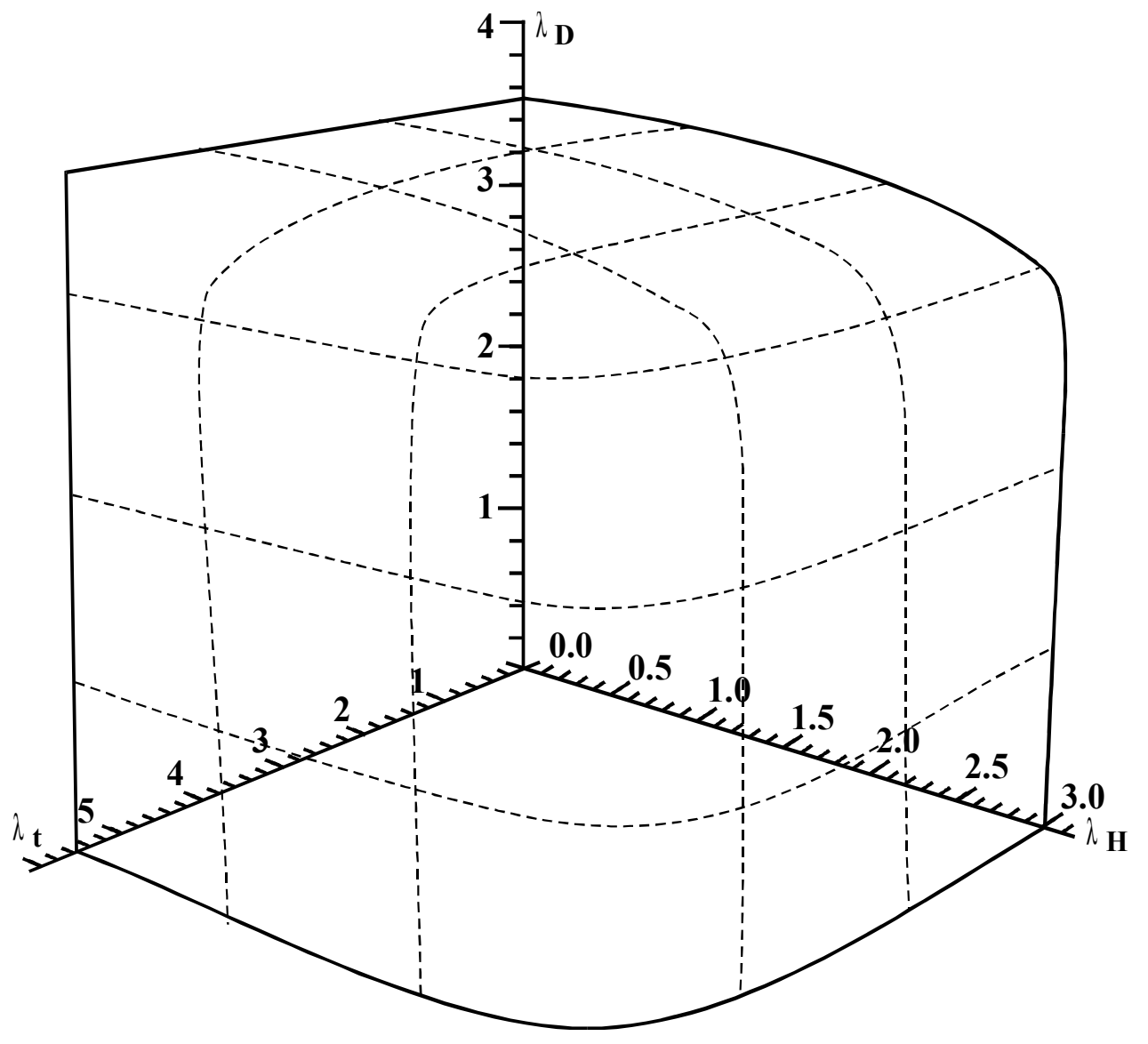

Fig. 4

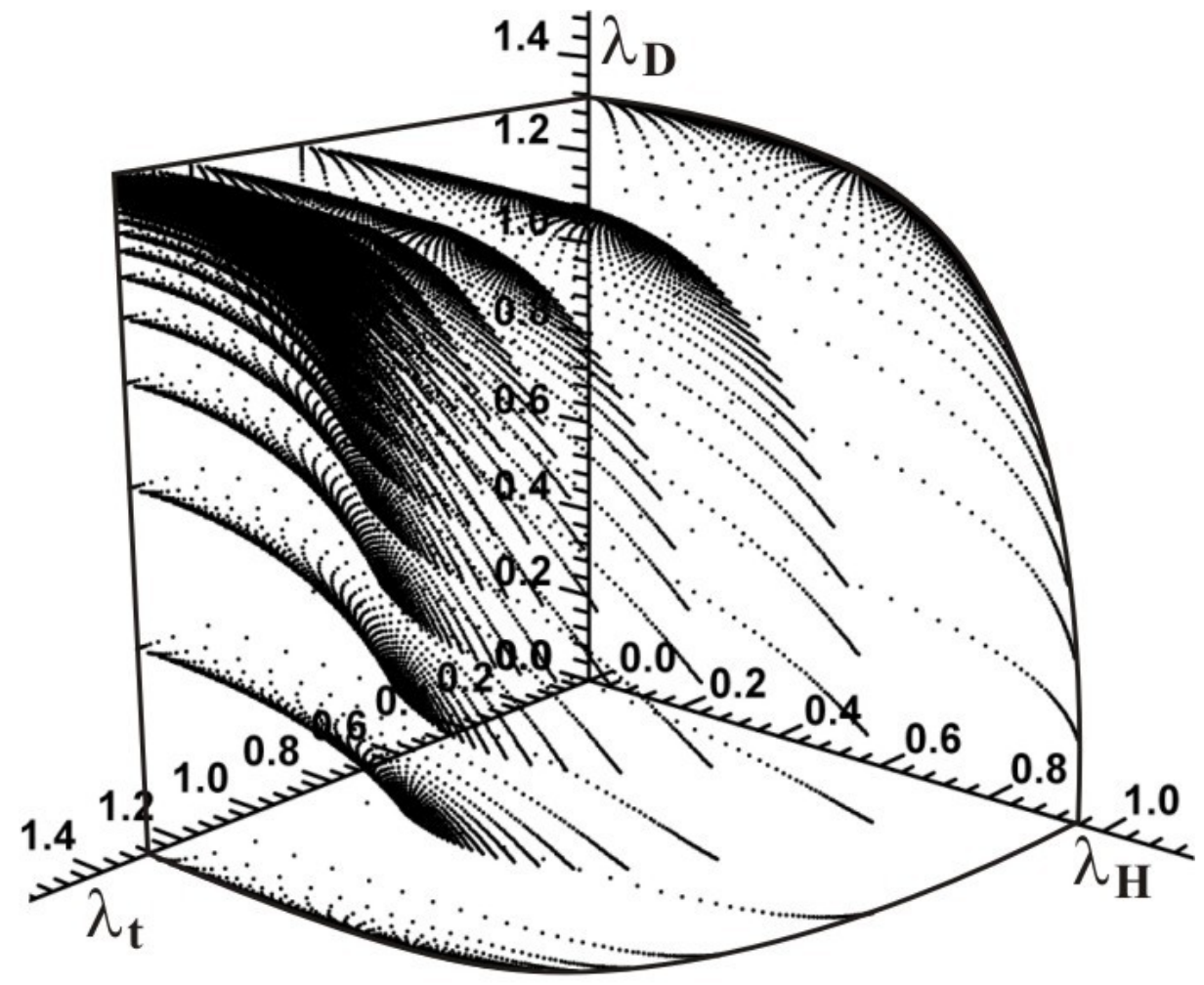

Fig. 5

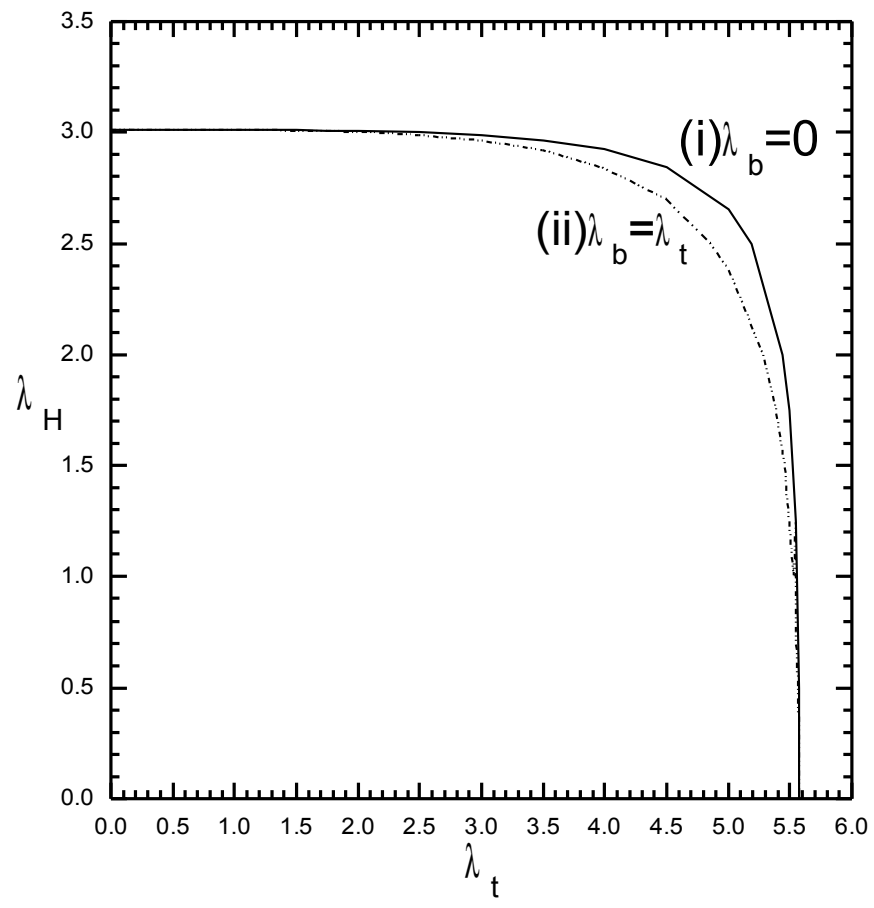

Fig.6(a)

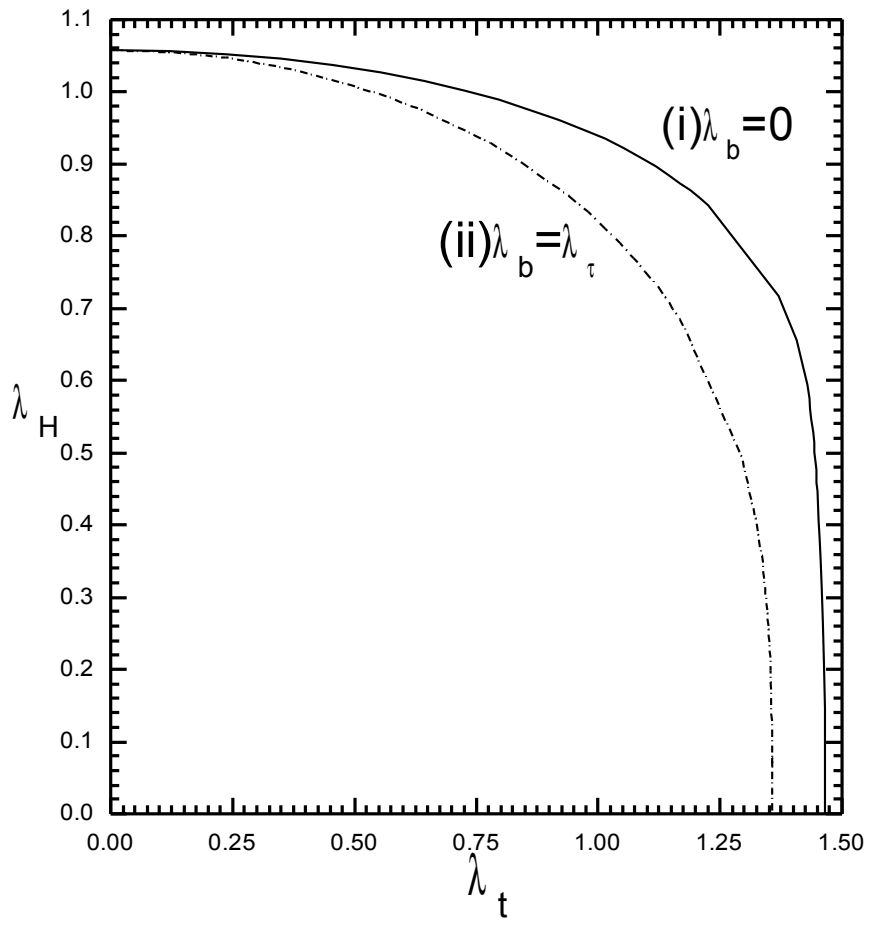

Fig.6(b)

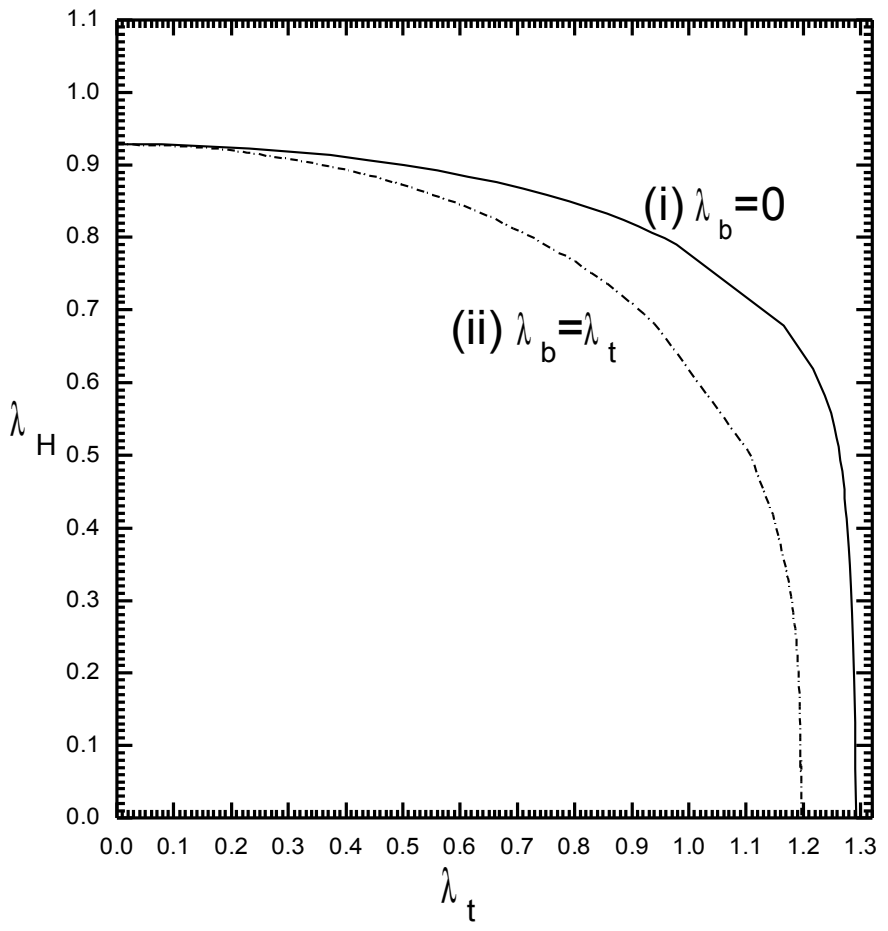

Fig.6(c)

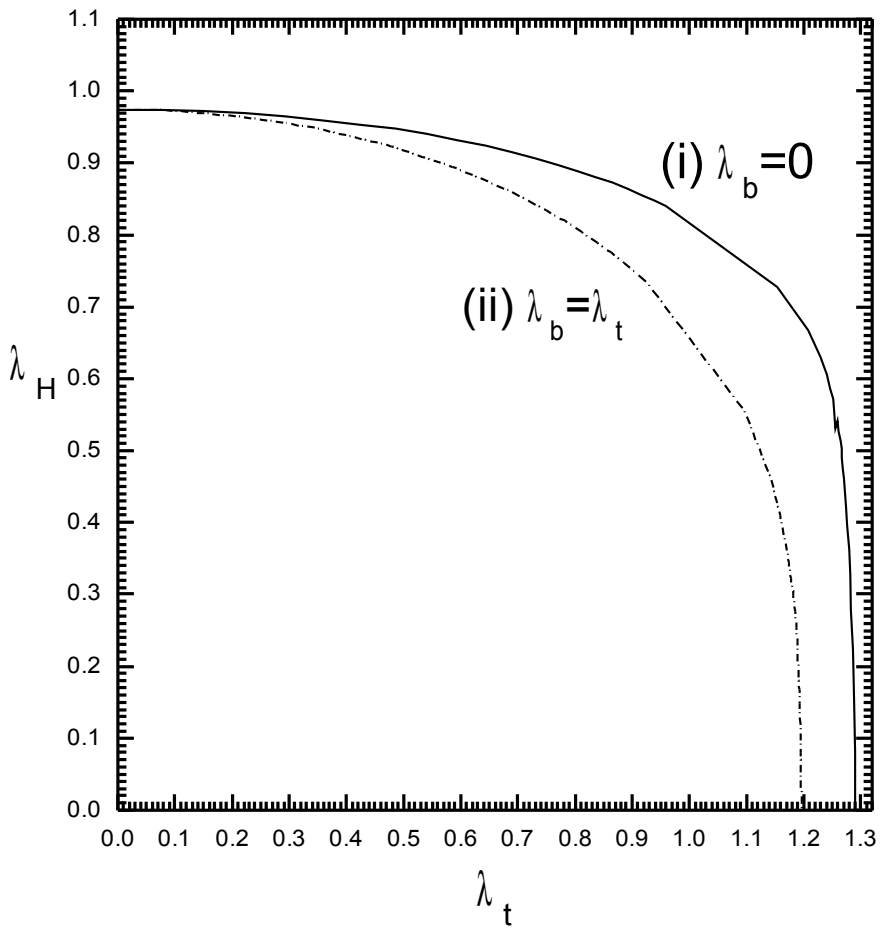

Fig.6(d)

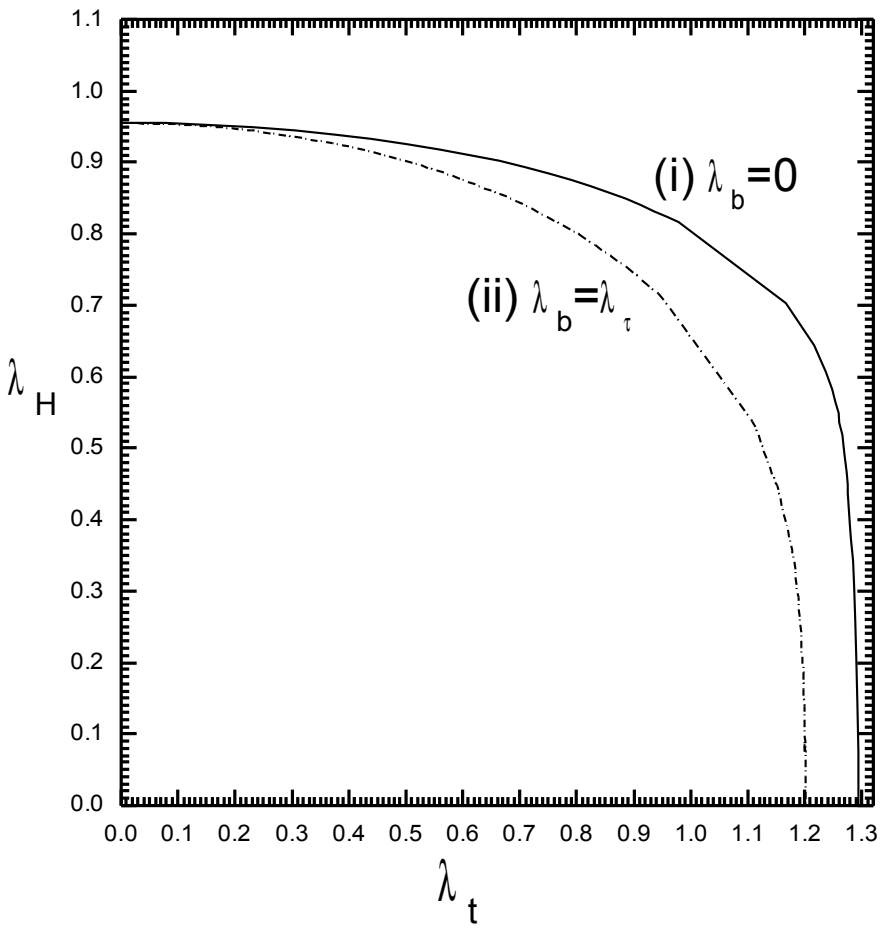

Fig.6(e)

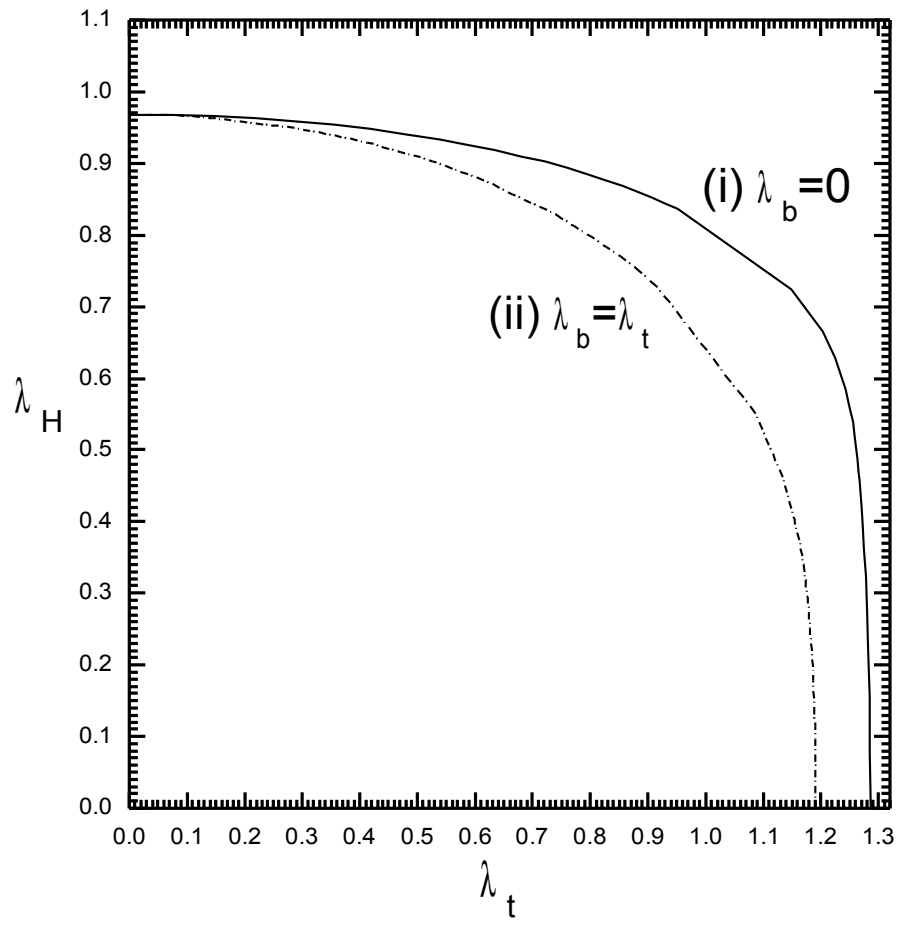

Fig.6(f)

**Fig. 6**

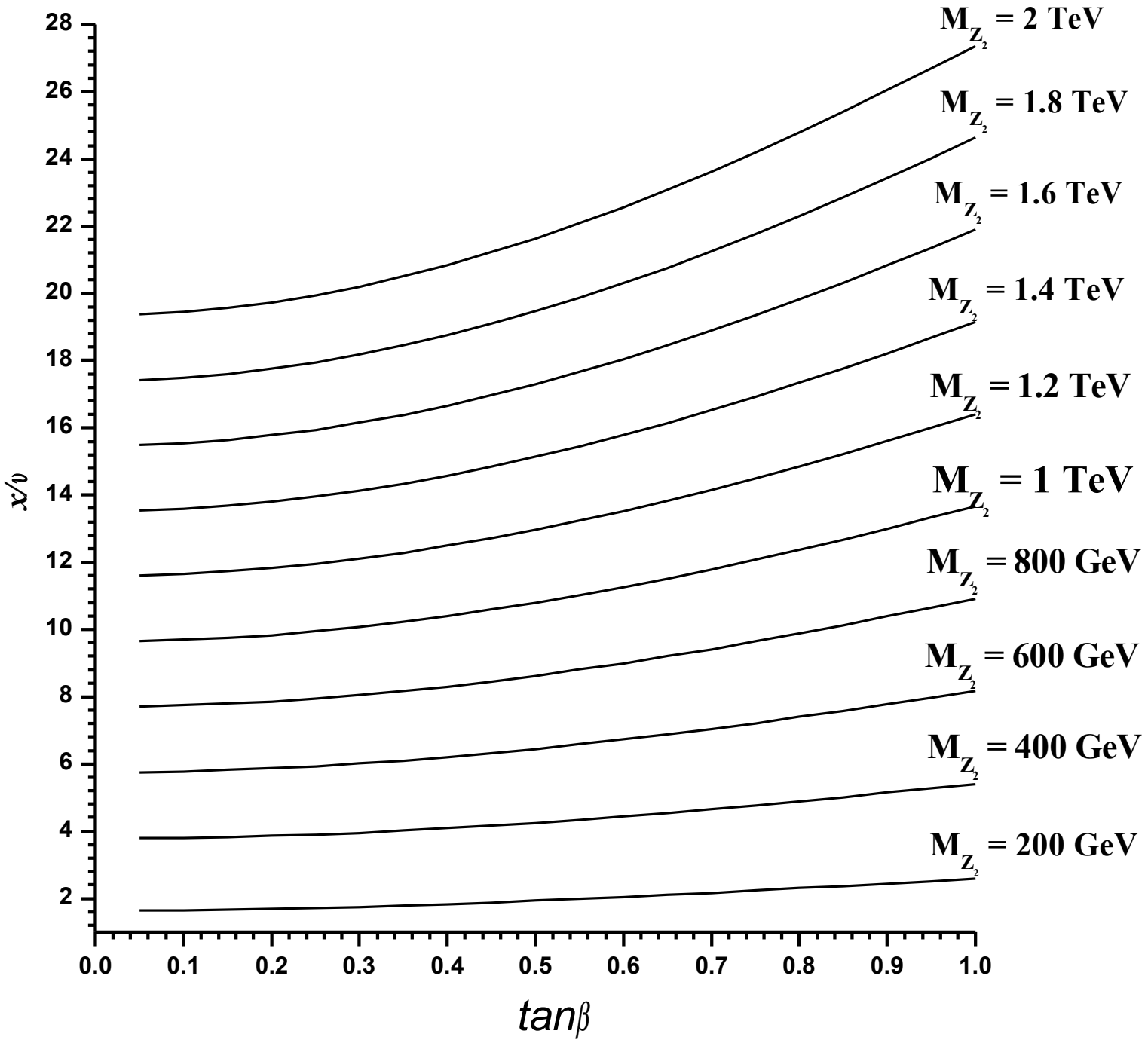

$M_{Z_2}$ = 2 TeV
$M_{Z_2}$ = 1.8 TeV
$M_{Z_2}$ = 1.6 TeV
$M_{Z_2}$ = 1.4 TeV
$M_{Z_2}$ = 1.2 TeV
$M_{Z_2}$ = 1 TeV
$M_{Z_2}$ = 800 GeV
$M_{Z_2}$ = 600 GeV
$M_{Z_2}$ = 400 GeV
$M_{Z_2}$ = 200 GeV
x/v
tanβ

Fig. 7(a)

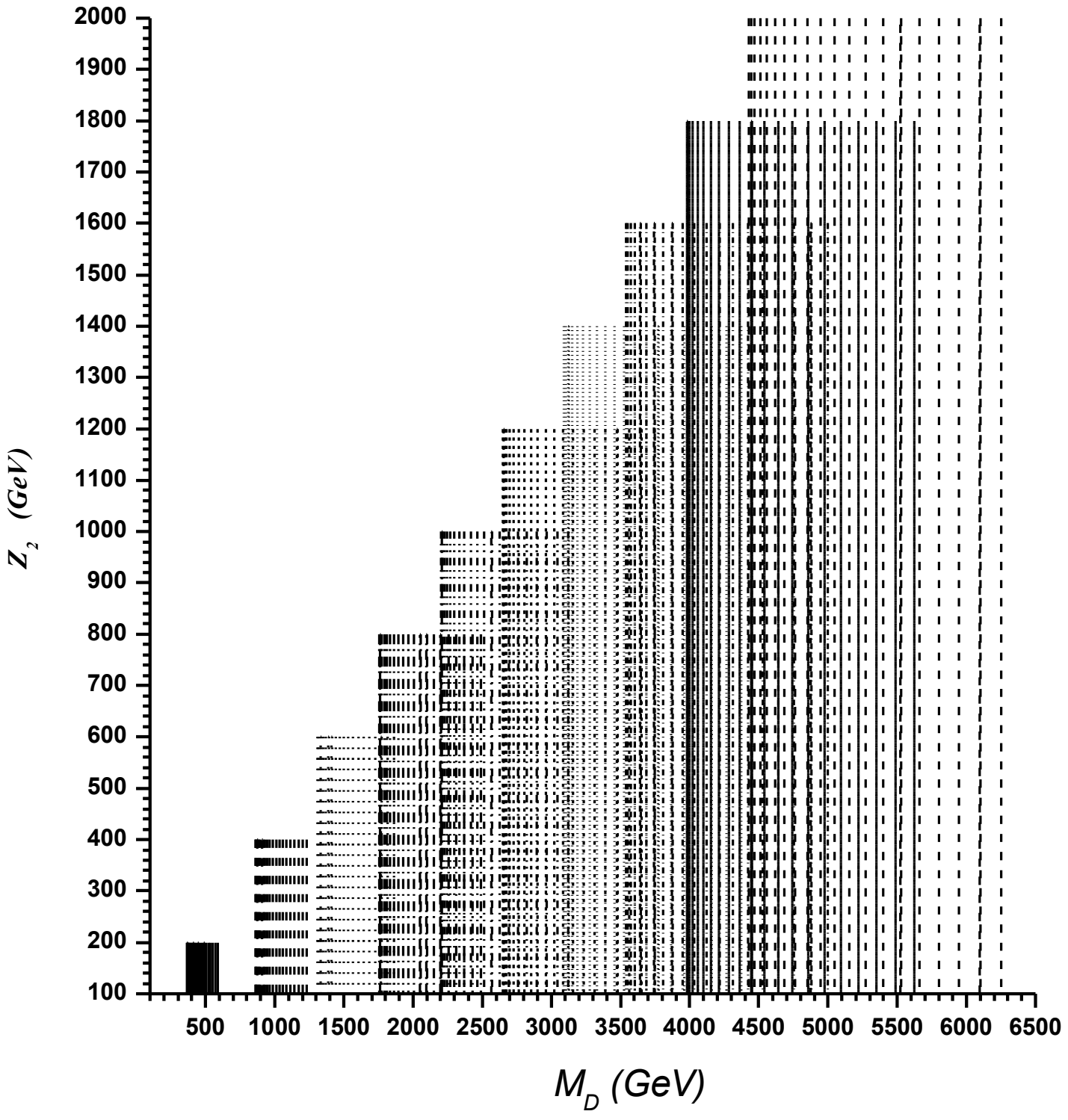

$Z_2$ (GeV)
$M_D$ (GeV)

Fig. 7(b)